\documentclass[final,times,11pt]{elsarticle}

\usepackage[
top    = 2.5cm,
bottom = 2.5cm,
left   = 2.5cm,
right  = 2.5cm]{geometry}

\usepackage{amssymb}
\usepackage{graphicx,amssymb,mathrsfs,epstopdf}
\usepackage{placeins}  
\usepackage{amsmath}
\usepackage{listings}
\usepackage{url}
\usepackage{xspace}
\usepackage{hyperref}
\usepackage{subfigure}
\usepackage{fancyvrb}
\usepackage{upquote}
\usepackage{varwidth}
\usepackage{float}
\usepackage{color}
\usepackage{rotating}
\usepackage{multirow}
\usepackage{textcomp}
\usepackage{array}
\usepackage{algorithm}

\makeatletter
\let\@afterindenttrue\@afterindentfalse
\makeatother

\providecommand{\tabularnewline}{\\}
\floatstyle{ruled}
\newfloat{algorithm}{tbp}{loa}
\providecommand{\algorithmname}{Algorithm}
\floatname{algorithm}{\protect\algorithmname}

\newenvironment{lyxlist}[1]
{\begin{list}{}
{\settowidth{\labelwidth}{#1}
 \setlength{\leftmargin}{\labelwidth}
 \addtolength{\leftmargin}{\labelsep}
 }}
{\end{list}}

\journal{Computer Physics Communications}

\begin{document}

\begin{frontmatter}



\title{\emph{CutLang:} A  Particle Physics Analysis Description Language and Runtime Interpreter}


%\author{}
%\address{}

\author[uci,ead]{G\"{o}khan~\"{U}nel}
\address[uci]{University of California at Irvine, Department of Physics and Astronomy, Irvine, USA}
\ead{gokhan.unel@cern.ch}

\author[knu, ead]{Sezen Sekmen}
\address[knu]{Kyungpook National University, Daegu, South Korea}
\ead{ssekmen@cern.ch}

\begin{abstract}
This note introduces \emph{CutLang}, a domain specific language that aims to provide a clear, human readable way to define analyses in high energy particle physics (HEP) along with an interpretation framework of that language.  A proof of principle (PoP) implementation of the \emph{CutLang} interpreter, achieved using C$^{++}$ as a layer over the CERN data analysis framework {\tt ROOT}, is presently available. This PoP implementation permits writing HEP analyses in an unobfuscated manner, as a set of commands in human readable text files, which are interpreted by the framework at runtime.
We describe the main features of \emph{CutLang} and illustrate its usage with two analysis examples.
Initial experience with \emph{CutLang} has shown that a just-in-time interpretation
of a human readable HEP specific language is a practical alternative to analysis writing using compiled languages such as C$^{++}$. \\

\end{abstract}

\begin{keyword}


LHC, data analysis, human readable language, runtime interpreter
\end{keyword}

\end{frontmatter}

\clearpage

\tableofcontents

\clearpage

\section{Introduction}

Since the era of the LEP experiments, particle physicists have been
performing large scale computing tasks to analyze particle physics data
in order to obtain physics results. A typical analysis requires extensive
manipulation of real and simulated particle collision events,  principally  
defining analysis objects, defining quantities that help
classify events as signal or background, selecting events, reweighting
simulated events to improve the agreement between the simulated and real events, and 
interpreting experimental results by comparing them to predictions.
Performing all of these tasks in a systematic manner generally requires
an analysis framework that organizes and sequences these tasks appropriately. 

As a consequence, a physicist who wishes to engage in a particle physics data
analysis needs to be well acquainted with computing at the level of
both a system programmer and a software developer. System level expertise is
needed because the frameworks generally comprise multiple software components,
while software level expertise is  needed in order to write these components. 
The coding needs
for analysis tasks include compiled languages like \texttt{C}, \texttt{C}$^{++}$
\cite{CPP}, and interpreted languages like \texttt{Python}~\cite{Rossum:1995:PRM:869369,Python}, \texttt{awk}~\cite{Aho:1987:APL:29361}
and \texttt{bash}~\cite{bash}. This list, combined with the structural
complexity and diversity in analysis framework design makes data analysis
a rather daunting task, and erects a barrier between data and the physicist who may
simply wish to try out an analysis idea. The complexities and technical difficulties have grown 
considerably
in the LHC era, both due to the unprecedented amount of data collected
by the LHC experiments, and to the increasingly elaborate analyses
inspired by these data. 

In recent years, these challenges have motivated the exploration of ways to
tame the complexity.
For example, LHC experiments are converging towards fewer
analysis frameworks that are developed and maintained
by dedicated experts. On the phenomenology side, frameworks
such as \texttt{CheckMate} \cite{CheckMate-1,CheckMate-2,CheckMate-3}
and \texttt{MadAnalysis} \cite{MadA-1,MadA-2,MadA-3,MadA-4} exist,
which implement a sizable number of LHC new physics search analyses ready
for use in reinterpretation studies. Similarly, \texttt{Rivet}
\cite{Rivet-1,Rivet-2} hosts a large repository of LHC analyses implemented
by the original analysts. Using centrally maintained frameworks to
implement and run analyses already helps to automate and reduce technical
burdens in analysis work. However, this approach is still severely biased
towards physicists with considerable coding skills. 

An alternative approach is to decouple the description of all physics
components of an analysis from the software framework. Analyses would
thus be fully and unambiguously described by a framework-independent
domain specific language. Previous attempts at defining such 
languages (based on the LHCO format) were made in \cite{LHCO1,LHCO2}.
More recently, a thorough open discussion was started at the Les Houches
PhysTeV 2015 workshop to define the elements and general structure
of a language that could be broadly used by LHC physicists
for describing analyses. An initial proposal, called the Les Houches
Analysis Description Accord (LHADA) was released to the LHC community~\cite{lhada}. 
LHADA is a domain specific language 
with a strict set of syntax rules and a limited number of operators. A LHADA description
of an analysis together with the associated
self-contained functions, encapsulating non-trivial variables and, or,
algorithms, provide a complete and unambiguous description of the analysis. 

A framework-independent analysis description language has many advantages.
First, it makes the writing of analyses significantly more 
accessible by eliminating coding complexities. Second, having
a standard, framework independent language allows analysis preservation beyond the lifetimes
of the experiments or analysis software, and facilitates the abstraction,
visualization, validation, combination, reproduction, interpretation
and overall communication of the contents of particle physics analyses. Third,  such a language
benefits not only the physicist working on an analysis within an experimental
collaboration, but also colleagues from other experiments and phenomenologists. 
However, like other computer languages, 
an analysis description language is just that: a description. 
In order to execute a LHADA described analysis, an interpreter (or compiler)
is needed to convert the description into instructions that can actually be executed,
perhaps in some framework.
The key point is that the mapping from the analysis
description language to executing code must be \emph{automatic} for the reasons
discussed above. 

In this paper, we introduce \emph{CutLang}, which is both an analysis description language and 
the name of its interpreter, much the same way that {\tt Python} is both a programming language and
the name of its interpreter.  The \emph{CutLang} language follows the LHADA principles, 
but currently uses a syntax different from that of LHADA, as firstly, since work on \emph{CutLang} had started earlier, 
and secondly, although a LHADA proposal exists, details of its syntax have not yet been fully finalized.  
The unique strength of the \emph{CutLang} interpreter lies in its capability to perform \emph{interpretation at runtime}, 
without the need for compilation.
Runtime interpretation allows making rapid changes in an analysis, and hence is a very practical feature especially in the phase of analysis design.  Therefore, this tool is intended primarily for analysis design by experimentalists and phenomenologists. 
%\textcolor{blue}{\emph{CutLang} can be used in collider experiments such as ATLAS or CMS since it can easily be adapted to read in event data from a variety of different formats. Currently  \emph{CutLang} is being actively used in a full fledged ATLAS exotics analysis, with about 200 systematics variations easily folded in due to its modular implementation.  Evidently, it is possible to completely define an analysis in a generic way, using human readable text files starting from the object selection up to the very last limit or discovery plot. Nevertheless, one should note that the sources of systematics variations, the methods to access those and the statistical analysis tools are usually experiment specific: no generic software tool can be used as an out of the box solution unless some serious effort is spent in defining various accords in those areas. 

\emph{CutLang} can be used in collider experiments such as ATLAS or CMS since it can easily be adapted to read in event data from a variety of different formats. Currently  \emph{CutLang} is being actively used in a full fledged ATLAS exotics analysis, and is capable of handling all object and event selection requirements.  
Another {\em raison d'\^etre} of \emph{CutLang} is to provide a tool for the phenomenology community for studies on new analysis ideas, new kinematic variables, or tests of sensitivity to future experiments such as HL-LHC or the FCC
\emph{CutLang} furthermore intends to serve physicists and physics enthusiasts using open data, who may not necessarily be expert programmers.
With its easy to learn human readable syntax, the authors hope that \emph{CutLang} would break the barrier between the collider data and the analysts.   Such a tool would undoubtedly increase the number of analysts and democratize access to the data collected using public funds.

In the remainder of this note, we describe \emph{CutLang} in detail and present several analyses examples written in \emph{CutLang} syntax together with their runtime interpretation  using the \emph{CutLang} interpreter. 
Section~\ref{sec:The-conceptual-description} lays out the \emph{CutLang}
concept and design principles, followed by Section \ref{sec:The-implementation-details},
which explains the implementation details. Section \ref{sec:examples}
illustrates the \emph{CutLang} language with two analysis examples.
Section \ref{sec:Processing-speed} discusses the runtime execution speed
of \emph{CutLang} versus that of the corresponding C$^{++}$
implementation. Our conclusions and the outlook are given in Section
\ref{sec:Conclusions-and-outlook}. A detailed user manual is given in the Appendix~\ref{sec:userman} and example implementations of two published ATLAS analyses are presented in Appendix~\ref{sec:furtherexamples}.

\section{Concept and design principles\label{sec:The-conceptual-description}}

\emph{CutLang} adheres to the first design principle of human readability
of the analysis description. It adopts a very simple syntax for analysis
objects, event selection criteria and histogram definitions, and collects
all resulting analysis description in a simple text file. 

There are (at least) two possible methods for converting a description written in a domain specific language into an
executable computer program: one either compiles and links the description, that is, the program; or
one interprets the description on the fly.
The first method requires a parser that recognizes the rules of the analysis
description language and automatically converts the analysis
description into a standard high level computer code (e.g. C$^{++}$), which would consequently be compiled, linked, 
and run. A well-known example which uses this procedure to compute
Feynman diagrams is \texttt{CompHEP} or \texttt{CalcHEP} \cite{Comphep,Calchep}.
Work is already in progress for writing parsers that convert an analysis
description based on the current LHADA proposal into executable code
which would run with several state of the art analysis frameworks
\cite{Brooijmans:2018xbu} (Section 23). 

In the second interpreter method, each component of the analysis description would
be evaluated at runtime without compilation. {\tt Python} is a well-known 
example of this approach\footnote{Another method, which combines the best of both methods, and
which is used in the ROOT6 framework
\cite{ROOT}, is just-in-time (JIT) compilation in which the interpreter invokes a compiler incrementally,
instruction by instruction, just when it is needed.}.

While both approaches have their relative merits, the interpreted
approach has the practical advantage for the user of bypassing the complications,  
and inherent slowness, of the compilation stage. 
In an interpreted analysis system, it is quick
to add new selection criteria, change their execution order,
or cancel any criterion by simply commenting it out. The same idea
holds for the addition, deletion and order change of histograms.
For these reasons, the interpretation approach is explored as the second
design principle for \emph{CutLang}. 

Furthermore, an analysis description language should allow the user to define new
objects such as a reconstructed Z boson, and its associated properties such as its mass or charge, or new variables such as hadronic transverse momentum, angular variables, etc.  These definitions would not only make the analysis description file
shorter but also much more human readable and understandable. Naturally,
such definitions should be available to both event selection and histogram
filling commands. Therefore, $CutLang$'s third design principle is
to allow users to define new objects and variables.

\section{The implementation details\label{sec:The-implementation-details}}

%\textcolor{blue}{It would be good to clarify whether when you refer to the CutLang analysis framework you mean only the CutLang interpreter or the interpreter plus something else. I think what you really mean is the CutLang interpreter. But clarification is needed.}

The concepts and design principles presented in the previous section
have been implemented in the \emph{CutLang} interpreter and analysis framework.
The software package consists of a single executable, the interpreter,
and at least two input files: a text file containing the analysis described in
\emph{CutLang} and a ROOT file containing the events to be
analyzed. The package, which  manipulates Lorentz vectors and
histograms, is written using custom C$^{++}$
classes and classes from ROOT. Therefore, the whole suite basically works in any modern
Unix-like environment. The following subsections describe the working
principles of \emph{CutLang} language and interpreter components. Further practical details about
installing and using the \emph{CutLang} analysis framework are given in the user manual provided
in Appendix \ref{sec:userman}.

%\textcolor{blue}{\subsection{A simple example of an analysis description using \emph{CutLang}} I think it would help the reader considerably to start with a line-numbered annotated example of CutLang before you delve into the detailed descriptions.}

\subsection{Input event format}

%\textcolor{blue}{analysis framework -> interpreter}

The \emph{CutLang} interpreter benefits from ROOT's {\tt TLorentzVector}
class to handle particles internally, and has extended it to incorporate
other particle properties such as charge and the ability to print
the particle in various formats.  From
this basic particle class, other specific classes for electrons, muons,
photons, jets \emph{etc.} are derived with their specific properties.
Such an abstraction of the physics object data is another critical design principle.
It makes possible to place a clean separation between 
the formats of the input objects (which can differ among experiments and between
different analyses, even analyses 
within the same experiment), and the objects upon which \emph{CutLang} operates.
This gives \emph{CutLang} the
flexibility to receive input in multiple formats and varying format
versions, which can be defined as different plugins.  Currently \emph{CutLang}
can work with the public ATLAS open data (\texttt{ATLASOD}) ntuple and CMS open
data (\texttt{CMSOD}) ntuple~\cite{CERN-OD, ATLAS-OD}, Delphes~\cite{deFavereau:2013fsa}, LHCO~\cite{LHCO1, LHCO2} and FCC~\cite{FCCEDM} event data formats in 
addition to its own data format called \texttt{LVL0}. New data formats can easily be incorporated as input by adding a new \texttt{C}$^{++}$ class and editing some of the available files, as described in Appendix~\ref{sec:newinputformat}. 
All currently applicable data are flat ntuples stored in ROOT files,
which can be chained together (using \texttt{TChain}). A further advantage
of using ROOT is the ability to run the analysis in a single or multi
processor environment such as a PROOF farm \cite{PROOF}. The interpreter
uses native \texttt{C}$^{++}$ libraries, however, conversion to XML is
also possible using ROOT's relevant library functions.

A carefully designed analysis description language is both framework-independent and
independent of the object types particular to an experiment. An analysis
developed by physicists in ATLAS may well use a particular set of ATLAS object types. But there
is no necessity for the system that 
executes an analysis description, be it an interpreter or a compiled executable program, to operate upon the ATLAS object types. The \emph{CutLang} interpreter operates on a set of standard, extensible, object types, which
are built from different input formats.  Ideally, such standard object types could be introduced and maintained by the ROOT Team at CERN.  A potential benefit of that support might be to  encourage the standardization
of analysis objects across experiments and analyses.

\subsection{Predefined particle types \label{sec:predef-particles}}

The basic particles (i.e. physics objects) and their properties are
already predefined in the \emph{CutLang} interpreter using its internal data format
\texttt{LVL0} and ready to be used by the analyst. Particles are sorted internally
by decreasing transverse momentum, and particle indices (starting
at zero) are marked with an underline (\texttt{\_}) character.  Sorting based on other criteria, e.g. energy, will be implemented in the next version. The basic particles are given in Table \ref{tab:Basic-physics-objects}. For example the
untagged jets are referred to with the keyword \texttt{JET}, the \emph{b}-tagged
jets with \texttt{BJET} and finally the light (quark gluon) jets with
\texttt{QGJET}~\footnote{Although the \emph{b}-tagging thresholds for different
input file types are defined with their 2017 values, it is possible
for the user to change this default by modifying the relevant variable
in the analysis description file. This variable is expected to be
written as \texttt{btag\_lowthreshold\_} followed by the input file
type, such as \texttt{ATLASOD} as listed in the previous subsection.}.
Missing transverse energy in the event is defined with the METLV keyword,
which is mapped to a Lorentz vector with zero axial momentum. As shown
in the last row of Table \ref{tab:Basic-physics-objects}, this particle
can only have 0 as its index value.

There are two object types that merit special attention, which are
the leptons and the neutrinos. First, for the leptons, the \texttt{LEP}
keyword is generic and can be reduced to refer either to an electron or a muon depending
on the trigger choice made in Table \ref{tab:Physics-objects-acceptances}.
This avoids the need to write two separate description blocks 
(see Section \ref{subsec:MultipleRegions}) for electron and muon
based selections in an analysis.  The trigger choice for a lepton channel either deselects it (when 0), 
treats it as data (when 1) or treats it as simulation applying all the Monte Carlo weights (when 2).
The second case is related to the treatment of neutrinos. At the LHC energies
and beyond, for which this tool is intended, the W bosons are generally
produced with a sufficient boost such that in the leptonic W decays,
the pseudorapidity of the charged lepton is not very different from that of
the neutrino. To address this case, \emph{CutLang} language defines a special
neutrino object. This object, denoted as \texttt{NUMET} is defined as a massless
and chargeless particle with transverse momentum and azimuthal angle
($\phi$) values extracted from the missing transverse energy, but
additionally, whose pseudorapidity is assumed equal to that of the
associated charged lepton. The lepton association is provided with
the, by now familiar, underline (\texttt{\_}) character.  
The possibility for the analyst to define new particle sets from existing sets, e.g. defining tight electrons from electrons, is also possible in \emph{CutLang}, and will be described in Section~\ref{sec:derivedobjectset}.

\begin{table}[h]
\caption{Default physics objects available in \emph{CutLang} \label{tab:Basic-physics-objects}}
\centering{}%
\begin{tabular}{|r|l|l|l|l|}
\hline 
Name & Keyword & Highest Pt object & Second Highest Pt object & $j^{th}$Highest Pt object\tabularnewline
\hline 
\hline 
electron & \texttt{ELE} & \texttt{ELE\_0} & \texttt{ELE\_1} & \texttt{ELE\_j}\tabularnewline
\hline 
muon & \texttt{MUO} & \texttt{MUO\_0} & \texttt{MUO\_1} & \texttt{MUO\_j}\tabularnewline
\hline 
lepton & \texttt{LEP} & \texttt{LEP\_0} & \texttt{LEP\_1} & \texttt{LEP\_j}\tabularnewline
\hline 
photon & \texttt{PHO} & \texttt{PHO\_0} & \texttt{PHO\_1} & \texttt{PHO\_j}\tabularnewline
\hline 
jet & \texttt{JET} & \texttt{JET\_0} & \texttt{JET\_1} & \texttt{JET\_j}\tabularnewline
\hline 
b-tagged Jet & \texttt{BJET} & \texttt{BJET\_0} & \texttt{BJET\_1} & \texttt{BJET\_j}\tabularnewline
\hline 
light Jet & \texttt{QGJET} & \texttt{QGJET\_0} & \texttt{QGJET\_1} & \texttt{QGJET\_j}\tabularnewline
\hline 
neutrino & \texttt{NUMET} & \texttt{NUMET\_0} & \texttt{NUMET\_1} & \texttt{NUMET\_j}\tabularnewline
\hline 
missing ET & \texttt{METLV} & \texttt{METLV\_0} & N/A & N/A\tabularnewline
\hline 
\end{tabular}
\end{table}

\subsection{Analysis description file \label{sec:inifile}}

The analysis description text file, commonly referred to as an \texttt{ini}
file, contains the \emph{CutLang} description of an analysis.  As
noted above, the \emph{CutLang} language differs from that of the LHADA proposal,
which is still being refined.  However, as also noted above, the \emph{CutLang} syntax is 
built on the same principles described within the LHADA proposal. Additionally, the
Extensible Markup Language (XML) format might also be used instead
of the \emph{CutLang} syntax, which may potentially simplify the editing process depending
on the editor used. 

There are three sections in the \texttt{ini} file: i) the physics
objects' acceptance values, ii) the derived object set and user variable definitions,  
and iii) the \emph{CutLang} selection and histogramming
commands. Throughout the \texttt{ini} file, user comments and explanations
are preceded by a hash (\#) symbol. Mass, energy and momentum are all
written in giga electron Volts (GeV) and angles in radians. The keywords
and variables are all case sensitive.

\paragraph*{Section i} The first section consists of a list of keywords describing object
properties and their associated values separated by an equal sign.
The full list of the keywords is given in Table \ref{tab:Physics-objects-acceptances}.
For example, an electron candidate should have a transverse momentum
above the given minimum value and an absolute value of pseudorapidity below the given maximum
to be accepted as a valid electron.  The last two lines in the same
table refer to the lepton (electron or muon) triggers. Monte Carlo
weights are not taken into account when the trigger value is set to
data. The threshold and trigger definitions in this section can be
done in random order.

\paragraph*{Section ii} The second section defines new objects or variables used in the analysis.
It includes definitions of new object sets derived from the default object sets in \emph{CutLang}.  
It also defines new particles and variables using object sets, and thus renders some commands clearer, for example, by introducing shortcuts like \texttt{Zhreco} for a hadronically reconstructed $Z$ boson. New particles and variables in this section can be defined in any order.
The derived object sets and the related keyword \texttt{obj} will be discussed in detail in Section~\ref{sec:derivedobjectset}, while defining new particles and variables along with the related keyword \texttt{def} will be discussed in Section~\ref{sec:particlevariable}. 

\paragraph*{Section iii} The third section defines the event manipulation commands, where a
command is either a selection criterion (where a selection criterion can be composed of multiple 
simpler criteria glued together by logical operators), or a special instruction to
include Monte Carlo weights, or to fill histograms. The command execution order is top to bottom.  
\emph{CutLang} language and interpreter permits the description and manipulation of
 multiple event selection regions.  Event selection and the related keywords \texttt{cmd} and \texttt{algo} will be discussed in Section~\ref{sec:evtselcmd}.  User can define histograms inside the event selection blocks, through a set of default histograms namely, the transverse momenta, $\eta$, $\phi$ and electric charge of all objects. 
Currently only one dimensional histograms are accepted.  Histogramming details will be discussed in Section\ref{sec:histos}.

\begin{table}
\caption{Physics objects acceptance thresholds and Trigger values\label{tab:Physics-objects-acceptances} }
\centering{}%
\begin{tabular}{|r|l|}
\hline 
Keyword & Explanation\tabularnewline
\hline  \hline 
\texttt{minpte} & minimum transverse momentum of electrons\tabularnewline \hline 
\texttt{minptm} & minimum transverse momentum of muons\tabularnewline \hline 
\texttt{minptg} & minimum transverse momentum of photons\tabularnewline \hline 
\texttt{minptj} & minimum transverse momentum of jets\tabularnewline \hline 
\texttt{maxetae} & maximum pseudorapidity of electrons \tabularnewline \hline 
\texttt{maxetam} & maximum pseudorapidity of muons\tabularnewline \hline 
\texttt{maxetam} & maximum pseudorapidity of photons\tabularnewline \hline 
\texttt{maxetaj} & maximum pseudorapidity of jets\tabularnewline \hline 
\texttt{TRGm} & 0=no trigger, 1=Data trigger for muons, 2=Monte Carlo trigger for muons\tabularnewline \hline 
\texttt{TRGe} & 0=no trigger, 1=Data trigger for electrons, 2=Monte Carlo trigger for electrons\tabularnewline \hline 
\end{tabular}
\end{table}

\vspace{1cm}

\subsection{Event selection commands \label{sec:evtselcmd}}
A selection class and various daughter classes inheriting from it
lie at the heart of the\emph{ CutLang} interpretation library. Each
selection class consists of a function that evaluates event by event
object or event-related quantities (e.g. $f(x)$ ), a comparison operator
(e.g. $>$), and one or two limit values. 
For each evaluation, the
function result is compared to the limit values to find a boolean
result of 0 or 1. Multiple selection criteria can be combined together
using two logical operators: \texttt{AND} and \texttt{OR}. Therefore,
each \emph{CutLang} selection command can be reduced to an arithmetical
expression consisting of additions and multiplications. Consider, for example,
the selection command
\begin{lstlisting}
nELE == 2 AND (( nJET > 2 ) OR ( nJET == 0)) AND nBJET == 0
\end{lstlisting}
For an event with two electrons and no jets, these functions are evaluated
to yield the arithmetic expression $1\times((0)+(1))\times1$ . The \emph{CutLang} interpreter 
evaluates such expressions numerically by first converting them into
Reverse Polish Notation (RPN) \cite{RPN} and then by parsing them
with the Shunting Yard algorithm \cite{ShuntingYard}. The resulting
boolean value determines whether the event is accepted or rejected.
The function value can also be extracted for histogramming purposes.
A typical command is given with the keyword \texttt{cmd}, followed
by a function, a comparator and limit value(s). The \emph{CutLang} syntax
requires the text after the \texttt{cmd} keyword to be enclosed
within quotation marks. One especially useful selection command is
ALL, which can be written as:
\begin{lstlisting}
cmd "ALL "
\end{lstlisting}
The purpose of this command is to select and count all events entering the 
event loop in order to later use them for selection efficiency calculation.
\emph{CutLang} automatically fills counting histograms after each step
in event selection. The remainder of this section lists the details of the object functions
and various comparison operators.

\subsubsection{Functions}

\emph{CutLang} currently has by default 14 simple functions with no
arguments and 12 functions that take particles as arguments. A full
listing of the functions with no arguments is shown and briefly explained
in Table \ref{tab:Operators-simplest}. These mostly consist of counting
functions and functions that do simple reconstructions such as the
scalar sum of the transverse momenta of all jets in an event (\texttt{SumHTJET}).
The list of the more complex functions requiring one or more particles
as arguments is given in Table \ref{tab:Functions}. In \emph{CutLang}
syntax, the particle argument list is given before the function
name. All such functions start with a right curly brace \texttt{\{}
and include commas to separate multiple particle arguments. An example
would be \texttt{\{ ELE\_0 \}q} which returns the charge for the electron
(or positron) with the highest transverse momentum. Another example
would be \texttt{\{ ELE\_0 , JET\_1 \}dR} , which computes the angular
distance ($R\equiv\sqrt{(\Delta \eta)^{2}+(\Delta\phi)^{2}}$) between the leading
electron and the sub-leading jet.\emph{ }

The function expressions like \texttt{\{ ELE\_0 \}q} or \texttt{\{
ELE\_0 , JET\_1 \}dR} can either be used directly in the event selection,
or alternatively be assigned to a variable name using the \texttt{def} keyword
prior to the event selection, in the variable definition section of
the \texttt{ini} file, as shown in the two examples below:
\begin{lstlisting}
def "ele0q : { ELE_0 }q"
def "dRelejet : { ELE_0 , JET_1 }dR"
\end{lstlisting}
The current \emph{CutLang} interpreter has a number of internally implemented functions
which can be intermixed using the four arithmetic operations and the power operator. For more complicated calculations,
the benefits of using experiment specific, externally defined functions is evident. This feature is 
available in the current version of  \emph{CutLang}.  Defining arithmetic operations and more complex functions will be discussed in detail in Section \ref{sec:userfunc}.
\begin{table}[h]
\caption{Simple (without argument) \emph{CutLang} functions. \label{tab:Operators-simplest}}
\centering{}%
\begin{tabular}{|r|>{\raggedright}m{0.55\columnwidth}|}
\hline 
Function & Returned quantity\tabularnewline \hline  \hline 
\texttt{nELE} & number of electrons\tabularnewline \hline 
\texttt{nMUO} & number of muons\tabularnewline \hline 
\texttt{nPHO} & number of photons\tabularnewline \hline 
\texttt{nLEP} & number of leptons {\footnotesize{}(electrons or muons, trigger dependent)}\tabularnewline \hline 
\texttt{nJET} & number of jets\tabularnewline \hline 
\texttt{nBJET} & number of \emph{b}-tagged jets\tabularnewline \hline 
\texttt{nQGJET} & number of light jets\tabularnewline \hline 
\texttt{HT} & sum of all the jets transverse momenta\tabularnewline \hline 
\texttt{METMWT} & sum of the leptonically reconstructed W boson's transverse mass and
missing transverse energy\tabularnewline \hline 
\texttt{MWT} & transverse mass of leptonically reconstructed \emph{W} boson \tabularnewline \hline 
\texttt{MET} & missing transverse energy\tabularnewline \hline 
\texttt{ALL} & all events\tabularnewline \hline 
\texttt{LEPsf} & inclusion of lepton MC scale factors\tabularnewline \hline 
\texttt{FillHistos} & filling histograms defined afterwards\tabularnewline \hline 
\end{tabular}
\end{table}

\begin{table}[h]
\caption{\emph{CutLang} functions with particle arguments\emph{.} \label{tab:Functions}}
\centering{}%
\begin{tabular}{|r|l|l|}
\hline 
Returned quantity & Function & Argument\tabularnewline \hline  \hline 
Mass of & \texttt{\{ \}m} & a particle\tabularnewline \hline 
Charge of & \texttt{\{ \}q} & a particle\tabularnewline \hline 
Phi of & \texttt{\{ \}Phi} & a particle\tabularnewline \hline 
Eta of & \texttt{\{ \}Eta} & a particle\tabularnewline \hline 
Absolute value of eta of & \texttt{\{ \}Eta} & a particle\tabularnewline \hline 
Transverse momentum of & \texttt{\{ \}Pt} & a particle\tabularnewline \hline 
Axial momentum of & \texttt{\{ \}Pz} & a particle\tabularnewline \hline 
Total momentum of & \texttt{\{ \}P} & a particle\tabularnewline \hline 
Energy of & \texttt{\{ \}E} & a particle\tabularnewline \hline 
Number of b-jets & \texttt{\{ \}nbf} & a list of particles\tabularnewline \hline 
Angular distance between & \texttt{\{ \}dR} & two comma separated particles\tabularnewline \hline 
Minimum azimuthal angle between & \texttt{\{ \}dPhi} & two comma separated particles\tabularnewline \hline 
%Angular distance between & \texttt{\{ \}dEta} & two comma separated particles\tabularnewline \hline 
\end{tabular}
\end{table}

\subsubsection{Comparison operators and thresholds\label{sec:compop}}

The \emph{CutLang} interpreter understands the basic mathematical expressions and
comparisons. The operators\texttt{ ==}, \texttt{!=}, \texttt{<=},
\texttt{<}, \texttt{>}, \texttt{>=}, and their \texttt{Fortran} counterparts
\texttt{EQ}, \texttt{NE}, \texttt{LE}, \texttt{LT}, \texttt{GT} and
\texttt{GE} are recognized and correctly interpreted. Additionally,
square braces \texttt{[]} and \texttt{][} are used to define inclusive or
exclusive ranges respectively.  Although similar results could be obtained
with mathematical and logical operators, the utilization of the range
operators shorten the commands and increase the readability. One should
also note that the upper and lower boundaries in both cases are inclusive.
The range comparators available in \emph{CutLang} are listed in Table \ref{tab:Range-comparators}.

\begin{table}[h]
\caption{Range comparators in \emph{CutLang}\label{tab:Range-comparators}}
\centering{}%
\begin{tabular}{|c|c|}
\hline 
Keywords & Explanation\tabularnewline
\hline 
\hline 
\texttt{{[}{]}} & include range between limit values\tabularnewline
\hline 
\texttt{{]}{[}} & exclude range between limit values\tabularnewline
\hline 
\end{tabular}
\end{table}

\subsubsection{Logical operators}

The two logical operators \texttt{AND} and \texttt{OR} (together with
their \texttt{C/C}$^{++}$ notational counterparts \texttt{\&\&} and
\texttt{||} ) can be used in the \emph{CutLang} language and interpreter to combine two or more
selection criteria and thereby help simplify the definition of complex selection
regions involving multiple selection variables. As in regular boolean
expressions, arithmetic operators have precedence over logical operators.
In case of doubt, parentheses can be used to explicitly define the
order of evaluation. The two examples below illustrate the implementation
of the logical operators to combine selection variables expressed as
functions (Section~\ref{sec:compop}) or user defined variables
(Section~\ref{sec:particlevariable}).
\begin{lstlisting}
nELE > 2 AND nELE <= 4 
nLEP EQ 2 AND (( mLL > 200 ) OR ( mLL LT 100 )) AND nJET < 9
\end{lstlisting}

\subsubsection{Ternary operator}
Analysis writing almost always involves conditional statements, therefore it is highly practical for an analysis description to allow expression of such statements.  \emph{CutLang} is capable of handling conditional expressions, which are incorporated through the so-called ternary operator commonly used in several general purpose programming languages.  
\emph{CutLang} adapts the \texttt{C/C}$^{++}$ notation for the ternary operation, with the \texttt{?} and \texttt{:} signs for condition and true/false separation, respectively.  A ternary operation is thus written as \emph{condition} \texttt{?} \emph{true case command} \texttt{:} \emph{false case command}.  A typical usage in an analysis with muons could be:
\begin{lstlisting}
cmd   " nMUO == 0 ? ALL : { MUO_0 }Pt < 10 "
\end{lstlisting} 
where the command with the ternary operator should be read as follows:  "if there are no muons in the event, accept the event;  otherwise the highest transverse momentum muon should be less than 10 GeV to accept the event ".
\emph{CutLang} also allows nested ternary operations with a maximum of two levels of operations (which contain three conditions). Generic syntax for the nested ternary operation is \emph{condition-0 ? ( condition-1 ? true : false ) : ( condition-2 ? true : false ) }.  The ternary operator is right associative, therefore nested ternary operators are evaluated from right to left, without using the parentheses.

\subsubsection{Comparison operators for $\chi^{2}$ minimization}

In an analysis with a multitude of objects of the same type, the analyst
could desire to search for the best combination defined by a certain
criterion. Typical examples would be to find the jet combination that
would yield the best W boson mass in fully hadronic $t\bar{t}$
reconstruction, or to find the two charged leptons that would result
in the best Z boson mass in leptonic Z reconstruction. Search
for the optimal combination of a given number of particles can be
expressed in the \emph{CutLang} language using two special comparison operators,
\texttt{$\sim=$} and \texttt{$!=$} . Given a certain variable $x$
with an optimal value $v$, the operators $\sim=$ and $!=$ are used
to calculate the particle combination that gives an $x$ value respectively
``closest to'' and ``farthest from'' $v$. These two operators
can be used to express $\chi^{2}$ like operations. Such a search
using $\chi^{2}$ minimization is written in \emph{CutLang} notation
using particles with negative indices from the very beginning of the
analysis. For example, the statement ``find two leptons with a combined
invariant mass closest to 90.1 GeV'' is expressed in the\emph{ CutLang}
notation as \texttt{\{ LEP\_-1 LEP\_-1 \}m $\sim$= 90.1} . In such
cases the \emph{CutLang} interpreter finds the optimal pair of such particles, and
stores them per event for possible later use. The analyst could later
refer to such a pair with their original index (\texttt{-1}) and subsequently
apply a different selection criteria such as requiring the invariant
mass of the lepton pair to be in a certain range (e.g. \texttt{\{
LEP\_-1 LEP\_-1 \}m {[}{]} 80 100 }), or to histogram some of the
properties of this particle pair. If another particle of the same
type (e.g. another pair of leptons to form say a $Z'$) is to be found,
it is necessary to use a different but still negative index value. 

\subsubsection{Working with different event selection regions\label{subsec:MultipleRegions}}

In a typical analysis, it is very common to work with multiple regions
defined based on different event selection criteria. For example,
an analysis can have multiple signal selection regions accompanied
by several control regions to isolate certain background processes,
or validation regions to validate various analysis procedures. The \emph{CutLang} language 
accommodates the simultaneous definition of multiple selection regions
in the\texttt{ini} file as independent blocks, with the help of the keyword
\texttt{algo}. Each selection region begins with the \texttt{algo}
keyword followed by the name assigned to the region. The region name
has to start and end with two underline characters ``\_\_'' for
correct interpretation, e.g. \texttt{algo \_\_preselection\_\_}, \texttt{algo
\_\_SR1\_\_}, \texttt{algo \_\_CR1\_\_}, etc. \emph{CutLang} also
allows nesting the selection criteria defining a region block within
another. For example, it is possible to call the preselection region
criteria within a signal selection region block by simply using the
name of the preselection region, as shown below:
\begin{lstlisting}
algo __SR1__
__preselection__
HT > 300
...
\end{lstlisting}
In this case, \texttt{SR1} is obtained by applying the cuts in the preselection
region followed by $H_{T}>300$. The \emph{CutLang} interpreter processes all signal
regions defined in the \texttt{ini} file and stores information on each region
in a separate directory in the output ROOT file.

\subsection{Derived object sets \label{sec:derivedobjectset}}
 
\emph{CutLang} permits definition of new object sets based on some selection criteria. Such a requirement may arise, for example, when cleaner and tighter objects are needed in an analysis.  A new object set is defined using the \texttt{obj} keyword, followed by the name of the new object set and the name of the root (input) object set.  A column sign (\texttt{:})  is used for separating the two names.  The name of the new object set has to start with the predefined particle names discussed before in Section~\ref{sec:predef-particles}; i.e. \texttt{JETclean} is valid but \texttt{cleanJETs} is not. The conditions used to define the derived object set are listed in the subsequent lines using the \texttt{cmd} keyword as described previously in Section~\ref{sec:inifile}.  Each line is another condition to be applied to the object set, possibly reducing the number of elements in the set. If a complete set of objects or particles are to be handled (looped over), the particle index normally used after the underline  \texttt{\_} character is omitted.  The example below defines a set of cleaned jets, named \texttt{JETclean}  based on regular jets. The applied condition is to calculate the angular distance between each jet and all electrons in the \texttt{ELE} set, and accept only those jets which are separated from any electron in the \texttt{ELE} set by at least 0.2.   
\begin{lstlisting}
obj  "JETclean : JET "
cmd  "{ JET_ , ELE_ }dR >= 0.2 "
\end{lstlisting}
A derived object set might also be based on a previously defined derived set, such as defining  \texttt{JETcleanest} based on  \texttt{JETclean} with additional requirements.

\subsection{User defined composite particles and related variables\label{sec:particlevariable}}
\emph{CutLang} allows the addition of
two or more physics objects or particles to create a composite
particle. Well known examples are adding two electrons to reconstruct
a Z boson, adding 3 jets to reconstruct a top quark, adding two
or more particles to form new physics particles like \emph{$Z^\prime$},
$W^\prime$ etc. One should note that in the\emph{ CutLang} syntax, physics
object addition is indicated using a space `` ",
without using a + or any similar signs. Just like ordinary particles
or objects, user defined composite particles can be used in functions
that take particle arguments to obtain user defined variables. 

As in the case of functions, user defined composite particles
and variables can be either used directly in the event selection,
or assigned names in the ``user defined variables'' section of the
\texttt{ini} file through the \texttt{def} keyword. The second method is more convenient
as it creates easily remembered notations for new particle names,
such as \texttt{Zreco} for a reconstructed Z boson, or new
properties such as \texttt{qLL} for the total charge of a lepton pair.
Such definitions are also expected to help the physicist easily remember
various particle reconstructions in an event. In \emph{CutLang} notation,
the new particle or variable is defined using the colon (\texttt{:})
symbol with the obvious caveat that the name used on the left side of
the symbol must be unique. Some self explanatory examples for
defining new composite particles and related variables are given below.

\begin{lstlisting}
def "mLL : { LEP_1 LEP_0 }m"
def "qLL : { LEP_1 LEP_0 }q"
def "Zreco : LEP_1 LEP_0" 
def "dR(LL,J0) : { Zreco , JET_0 }dR" 
\end{lstlisting}

\subsection{User defined selection functions \label{sec:userfunc}}

Analyses often require computation of object or event quantities derived from predefined quantities.  
\emph{CutLang} is equipped to perform such computations in two different ways, depending on the complexity of the operations.  
Simple computations consisting only of basic arithmetic operations can explicitly be written into the analysis description file, and directly be executed by the \emph{CutLang} interpreter.  For example, a selection based on the quantity MET${/\sqrt{\mathrm{HT}}}$ would be written using the following syntax:

\begin{lstlisting}
cmd   " MET / HT ^ 0.5 > 20 "
\end{lstlisting}
If, as in the above example, the computation involves a predefined \emph{CutLang} function, it is possible to compute that function with a user defined derived object set (introduced in Section~\ref{sec:derivedobjectset}).  For example, HT, which originally uses \texttt{JET}, can be calculated with a user defined cleaned jets set named \texttt{JETclean}.  In this case, the previous command line should be modified as: 
\begin{lstlisting}
cmd   " MET / HT ^ 0.5 ( JETclean_ )  > 20 "
\end{lstlisting}
Arithmetic operations between predefined functions along with the ternary operator will address a substantial variety of cases, but modern realistic analyses occasionally require computation of more complex quantities.  The so-called ``external" functions are often used to implement such algorithms, and once defined for one analysis, are usually published for generic use.  However since the variable names and data access methods would be framework specific, such external functions can never be automatically incorporated nor used without a minimal editing and adaption, which could either be applied directly to the function itself, or implemented through an external adapter mechanism.  Currently, \emph{CutLang} chooses to take the former approach, and adapt the functions themselves. Therefore, a more suitable name for such functions is, in fact, ``user" functions.  \emph{CutLang} provides the means to automatically define and use a user function via a Python helper script. This script allows addition ( and deletion ) of a user function with a given name and its addition into the list of available selection functions. The user is only expected to write the main body of the selection function in C$^{++}$ and compile it in.  Further details on the implementation are given in Appendix~\ref{sec:userfuncdetail}.

\subsection{Histogram commands \label{sec:histos}}

Histograms to be filled at a particular analysis step can be straightforwardly
defined at the relevant step in the \emph{CutLang} \texttt{ini} file using
the \texttt{histo} keyword. Each histogram is defined by a number
of parameters. These are the histogram name that appears in the output
ROOT file, the histogram title, the number of bins, the minimum and
maximum values of the x-axis, and finally the name of the variable
to be histogrammed in the \emph{CutLang} notation. Any of the previously
defined user keywords or the predefined ones can be used in this comma
delimited set. An example using the above discussed definitions would
be:
\begin{lstlisting}
histo "Zlm , Leptonic Z reconstructed mass (GeV), 50, 50, 150, mLL " 
\end{lstlisting}
Currently only 1D histogramming is implemented. Additionally, there
are some predefined sets of histograms collectively denoted by the
keyword \texttt{Basics}. These are simply all the basic kinematic
distributions of physics objects (jets, electrons, muons etc.) such
as the pseudorapidity, transverse momentum, energy etc. distributions.
These basic histograms can be called using the command:
\begin{lstlisting}
histo "Basics "
\end{lstlisting}

Finally, one should note that any list of histogram definition commands
should always be preceded with a selection command \texttt{FillHistos} in order to 
ensure filling histograms that are defined afterwards. 

\subsection{\emph{CutLang} output}

The current \emph{CutLang} interpreter starts by listing
all selection criteria and histogram definition commands on screen
to prove its correct interpretation. At this initialization stage,
any unrecognized keywords are shown as ``\textbf{UFO}''s and presented
to the analyst as errors. During the event loop, the program prints
out the processed number of events with some very low frequency adjustable
by the user. 

At the final stage, the selection efficiencies of all selection regions
are printed and the resulting histograms are saved into ROOT files,
one for each region. Although these files can be used to display histograms
in various formats, a simple script exists for displaying those which
adhere to the simple convention of keeping the same histogram base
name followed by an integer showing the histogram's order \cite{showAllMacro}.

\section{Some analysis examples\label{sec:examples}}

In the following, two simple examples that illustrate the usage of
\emph{CutLang} for writing two generic LHC analysis flows involving
Z boson and top quark mass reconstruction are presented. 
Further examples are presented in Appendix~\ref{sec:furtherexamples}

\subsection{Z boson reconstruction}

The first simple example involves the reconstruction of the Z boson
mass from two charged leptons. The simple analysis algorithm written
in \emph{CutLang} notation is listed in Algorithm \ref{alg:Simple-Z-boson}.
The \texttt{ini} file containing this algorithm is run on MC events
in ATLAS and CMS open data ntuple format obtained from the CERN open
data portal \cite{CERN-OD}. 

The first section of the analysis description file shows threshold
values for various particles to define their detector acceptances.
The trigger is set to MC type electrons, meaning that in the lines
following the trigger selection, all particles indicated with the 
keyword \texttt{LEP} will be treated as electrons. Then comes the 
user defined variables section, where the reconstructed Z boson,
\texttt{Zreco} is defined using two leptons as a new particle, followed
by the definitions of the dilepton invariant mass \texttt{mLL} and
dilepton total electric charge \texttt{qLL}.

\begin{figure}
\begin{centering}
\includegraphics[width=0.43\columnwidth]{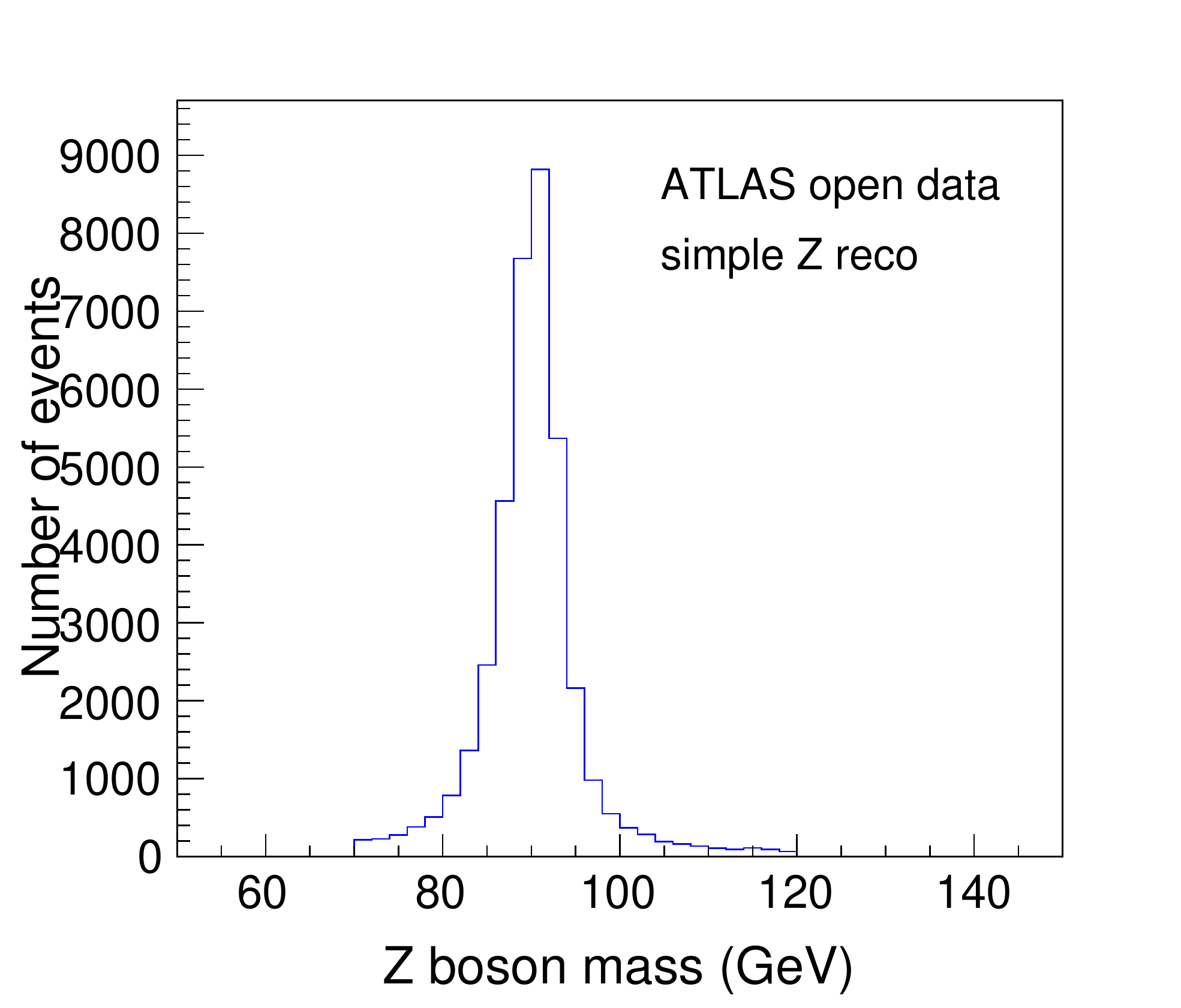}\includegraphics[width=0.43\columnwidth]{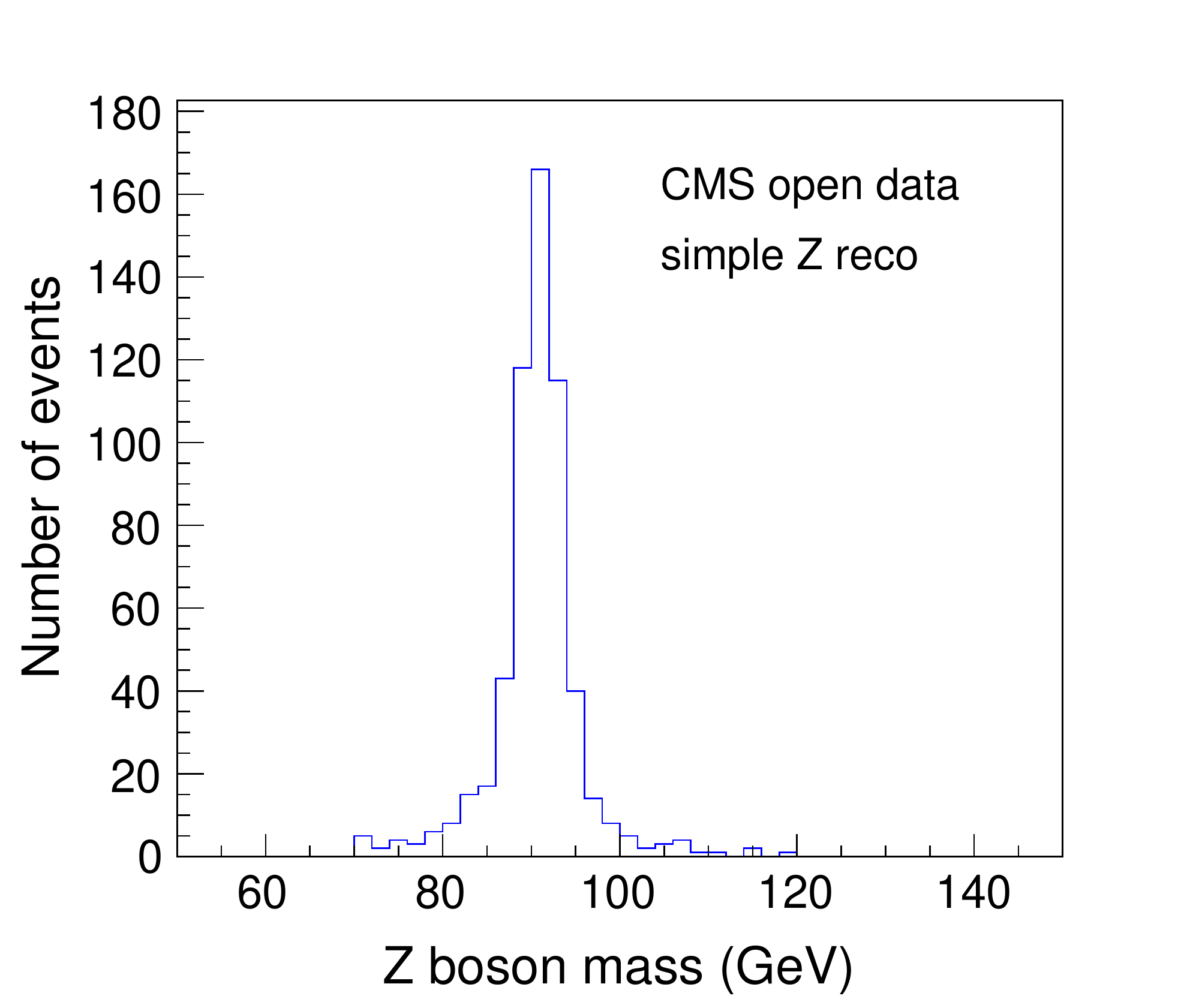}
\par\end{centering}
\caption{Dilepton invariant mass obtained from simple reconstruction algorithm.
ATLAS dielectron events are shown on the left and CMS dimuon events on
the right. \label{fig:Z-boson-simple}}
\end{figure}

The next section contains the event selection for a single signal
region. The first command \texttt{ALL} accounts for the total event
count by selecting all events, in order to later use this count in
the selection efficiency calculation. Next, two leptons are required,
which are needed to reconstruct the Z bosons. The following
selection on \texttt{qLL} ensures that these two leptons are oppositely
charged. Among the reconstructed Z boson candidates, only those in
the mass window of 70 to 120 GeV are selected. Finally the invariant
masses of the surviving lepton pairs are filled in histograms, which
are shown in Figure \ref{fig:Z-boson-simple} for ATLAS dielectron and CMS dimuon events.

\begin{algorithm}
\caption{Z boson reconstruction algorithm, case 1\label{alg:Simple-Z-boson}}
\begin{lstlisting}

###### PARTICLE THRESHOLDS
minpte  = 15.0  # min pt of electrons 
minptm  = 15.0  # min pt of muons 
minptj  = 15.0  # min pt of jets
maxetae = 2.47  # max pseudorapidity of electrons  
maxetam = 2.5   # max pseudorapidity of muons 
maxetaj = 5.5   # max pseudorapidity of jets

TRGm = 0 #     muon Trigger Type: 0=dont trigger, 1=1st trigger (data) 2=2nd trigger (MC)
TRGe = 2 # electron Trigger Type: 0=dont trigger, 1=1st trigger (data) 2=2nd trigger (MC)

###### USER DEFINITIONS
def       "mLL : { LEP_1 LEP_0 }m"
def       "qLL : { Zreco }q"         #note the nested definition utilization
def       "Zreco : LEP_0 LEP_1 "

###### EVENT SELECTION
cmd      "ALL "           # to count all events
cmd      "nLEP == 2 "     # events with only leptons
cmd      "qLL == 0 "      # reconstructed object should be neutral
cmd      "mLL [] 70 120 " # central mass for Z candidate
cmd      "FillHistos "
histo    "Zlm , Leptonic Zreco best (GeV), 50, 50, 150,  mLL  "
\end{lstlisting}
\end{algorithm}

\begin{figure}
\begin{centering}
\includegraphics[width=0.43\columnwidth]{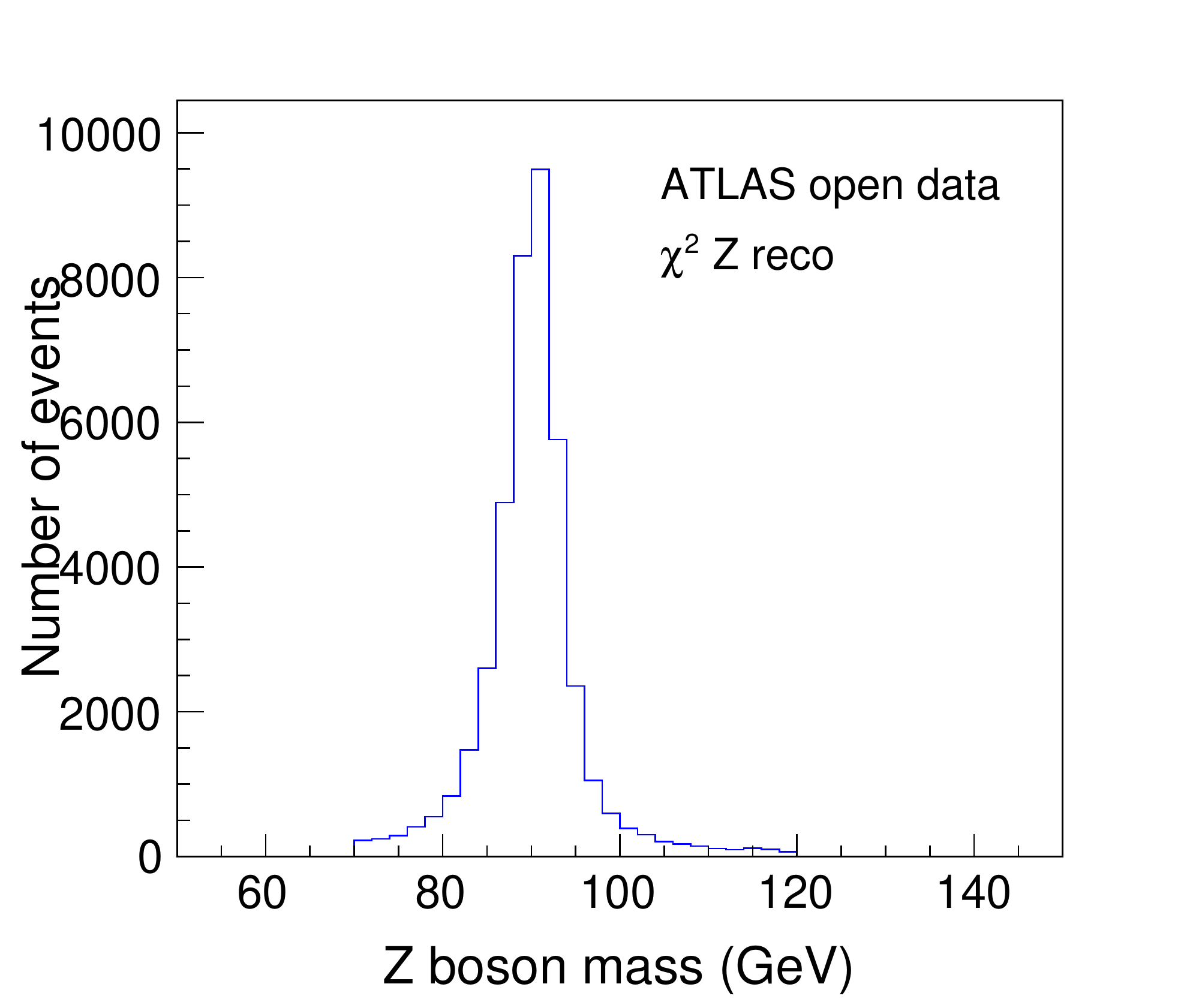}\includegraphics[width=0.43\columnwidth]{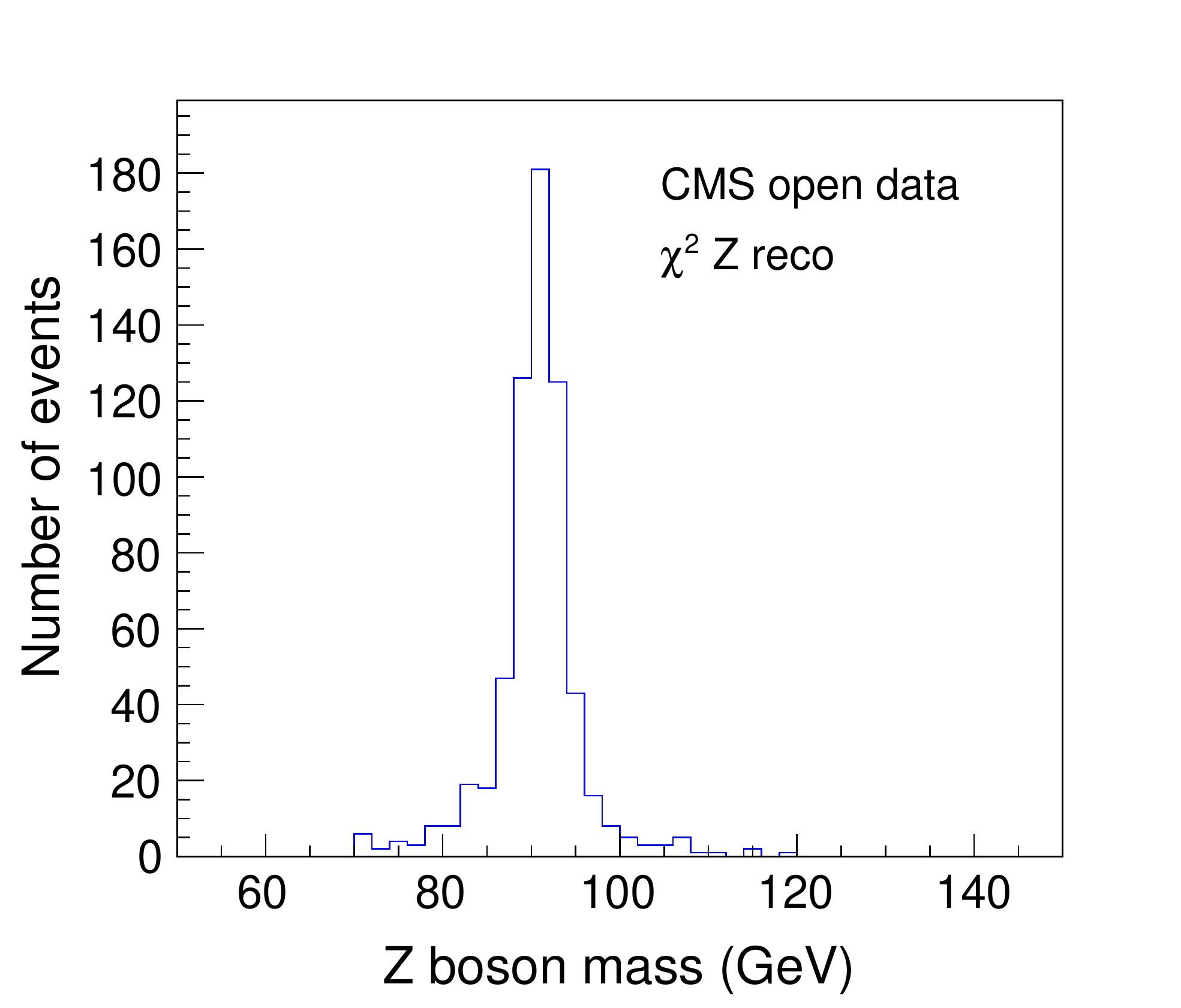}
\par\end{centering}
\caption{Z boson reconstruction example with automatic search. ATLAS dielectron
events are shown on the left and CMS dimuon events on the right.\label{fig:Zreco2}}
\end{figure}

A modified version of the above reconstruction is shown in Algorithm
\ref{alg:Zreco2} which illustrates the utilization of the $\chi^{2}$
search method. This algorithm selects events with two or more leptons.
It then reconstructs the Z boson from the lepton pair that
would give an invariant mass closest to that of the Z boson. The
``closest to'' operation itself does not select or reject events.
Selection is subsequently applied by requiring opposite sign leptons
and by defining an invariant mass window. Resulting \emph{Z} boson
invariant mass histograms are shown in Figure \ref{fig:Zreco2}.

\begin{algorithm}
\caption{Z boson reconstruction algorithm, case 2\label{alg:Zreco2}}
\begin{lstlisting}

###### PARTICLE THRESHOLDS
minpte = 15.0 # min pt of electrons 
minptm = 15.0 # min pt of muons 
minptj = 15.0 # min pt of jets
maxetae = 2.47 # max pseudorapidity of electrons 
maxetam = 2.5 # max pseudorapidity of muons 
maxetaj = 5.5 # max pseudorapidity of jets

TRGm = 0 # muon Trigger Type: 0=dont trigger, 1=1st trigger (data) 2=2nd trigger (MC)
TRGe = 2 # electron Trigger Type: 0=dont trigger, 1=1st trigger (data) 2=2nd trigger (MC)

###### USER DEFINITIONS
def       "mLL : { LEP_-1 LEP_-1 }m"
def       "qLL : { LEP_-1 LEP_-1 }q"

###### EVENT SELECTION
cmd      "ALL " # to count all events
cmd      "nLEP >= 2 " # events with only leptons
cmd      "mLL ~= 91.2 " # central mass for Z candidate
cmd      "qLL == 0 " # reconstructed object should be neutral
cmd      "mLL [] 70 120 " # central mass for Z candidate
cmd      "FillHistos "
histo    "Zlm , Leptonic Zreco best (GeV), 50, 50, 150, mLL "
\end{lstlisting}
\end{algorithm}

\subsection{Top quark pair reconstruction\label{subsec:ana-ttbar}}

This example illustrates two cases of top quark mass reconstruction from
three jets without using any $b$-tagging information. The first case
shows a two step top quark pair reconstruction where the first step
finds two W bosons and the second step combines the W bosons
with two jets to reconstruct top quarks. The second case tackles
the same problem in a single step, and directly finds the best three-jet 
combination that is most consistent with a top quark in terms of mass. The first case 
does a less accurate top reconstruction, but it is also less demanding from the computational point of view, and it
is a good illustration for a search algorithm in \emph{CutLang}. On the other
hand, one step reconstruction offers a more efficient top reconstruction
algorithm, but it is more computationally demanding due to the many
loops. This case represents a complex $\chi^2$ scenario for \emph{CutLang} that
is worth exploring. In both examples, fully hadronic events in ATLAS
Open Data format, downloaded from ATLAS public internet pages were
used. 

\subsubsection{Two step top quark pair reconstruction }

In this algorithm, after requiring at least 6 jets and relatively
low missing transverse energy, initially two W bosons are reconstructed
followed by combining the W boson candidates with the remaining
jets to form two top quarks according to the $\chi^{2}$ definition
in Equations \ref{eq:chi2wwtt} below: 

\begin{eqnarray}
\chi_{2w}^{2} & = & \frac{(m_{j_{1}j_{2}}-m_{W}^{MC})^{2}}{\sigma_{\Delta m_{W}^{MC}}^{2}}+\frac{(m_{j_{3}j_{4}}-m_{W}^{MC})^{2}}{\sigma_{\Delta m_{W}^{MC}}^{2}} \label{eq:chi2wwtt} \\
\nonumber \chi_{2t}^{2} & = & \frac{(m_{b_{1}j_{1}j_{2}}-m_{b_{2}j_{3}j_{4}})^{2}}{\sigma_{\Delta m_{bjj}}^{2}} 
\end{eqnarray}
Here $\Delta_{m_{w}^{MC}}$ is the experimentally expected width of
the W boson invariant mass distribution and $m_{W}^{MC}$ is the
expected W boson mass. In the first step, with a loop over
all jets, two separate jet pairs which would yield an invariant mass
as close as possible to W boson mass are selected. In the second
step, these reconstructed W boson pairs are used to define
the top quark candidates with the constraint of having the two reconstructed
objects' masses to be as close as possible to each other. In the second
equation, although the jets are denoted as $b_{1}$ and $b_{2}$,
no $b$-tagging is applied. For rendering the test more realistic
and also to demonstrate the abilities of \emph{CutLang}, an additional
selection is applied on the angular separation between the W 
boson and its associated jet forming the top quark. A total of 8 histograms
are filled with the analysis variables at various selection steps
to demonstrate usage of histograms. The full analysis algorithm description
in \emph{CutLang} syntax can be seen in Algorithm \ref{alg:wwtt-algo}.

\begin{algorithm}
\caption{Top quark pair two step reconstruction example with $\chi^{2}$ search
and without $b$-tagging.\label{alg:wwtt-algo}}
\begin{lstlisting}

###### PARTICLE THRESHOLDS
minpte  = 15.0  # min pt of electrons 
minptm  = 15.0  # min pt of muons 
minptj  = 15.0  # min pt of jets
maxetae = 2.47  # max pseudorapidity of electrons  
maxetam = 2.5   # max pseudorapidity of muons 
maxetaj = 5.5   # max pseudorapidity of jets

TRGm = 0 #     muon Trigger Type: 0=dont trigger, 1=1st trigger (data) 2=2nd trigger (MC)
TRGe = 2 # electron Trigger Type: 0=dont trigger, 1=1st trigger (data) 2=2nd trigger (MC)

###### USER DEFINITIONS
def       "WH1 : JET_-1 JET_-1 "  # W boson of the first top
def       "WH2 : JET_-11 JET_-11 "  # W boson of the second top
def       "mWH1 : { WH1 }m "  # mass of W boson of the first top
def       "mWH2 : { WH2 }m "  # mass of @ boson of the second top
def       "mTopH1 : { WH1 JET_-2 }m "  # first top quark's mass
def       "mTopH2 : { WH2 JET_-4 }m "  # second top quark's mass
def       "WHbR1 : {WH1 , JET_-2 }dR "  # angular distance between W1 and b jet
def       "WHbR2 : {WH2 , JET_-4 }dR "  # angular distance between W2 and b jet

###### EVENT SELECTION
cmd      "ALL "           # to count all events
cmd      "nJET >= 6 "     # events with 6 or more jets
cmd      "MET < 100 "     # fully hadronic events should have small MET
cmd      "mWH1 - 80.4 / 2.1 ^ 2 + mWH2 - 80.4 / 2.1 ^ 2 $\sim$= 0 "  # find 2 hadronic Ws
cmd      "( ( mTopH1 - mTopH2 ) / 4.2  )^ 2 $\sim$= 0 "
cmd      "FillHistos "
histo    "mWHh1 , Hadronic W best reco (GeV), 50, 50, 150, mWH1 "
histo    "mWHh2 , Hadronic W best reco (GeV), 50, 50, 150, mWH2 "
histo    "mTopHha1 , Hadronic top reco (GeV), 70, 0, 700, mTopH1 "
histo    "mTopHhb1 , Hadronic top reco (GeV), 70, 0, 700, mTopH2 "
histo    "WHbRh1 , Angular distance between W and bjet, 70, 0, 7, WHbR1 "
histo    "WHbRh2 , Angular distance between W and bjet, 70, 0, 7, WHbR2 "
cmd      "WHbR1 > 0.6 "
cmd      "WHbR2 > 0.6 "
cmd      "FillHistos "
histo    "mTopHha2, Hadronic top reco (GeV) after angular cut, 70, 0, 700, mTopH1 "
histo    "mTopHhb2, Hadronic top reco (GeV) after angular cut, 70, 0, 700, mTopH2 "
\end{lstlisting}
\end{algorithm}

\subsubsection{Single step top quark pair reconstruction}

\begin{figure}
\begin{centering}
\includegraphics[width=0.85\textwidth]{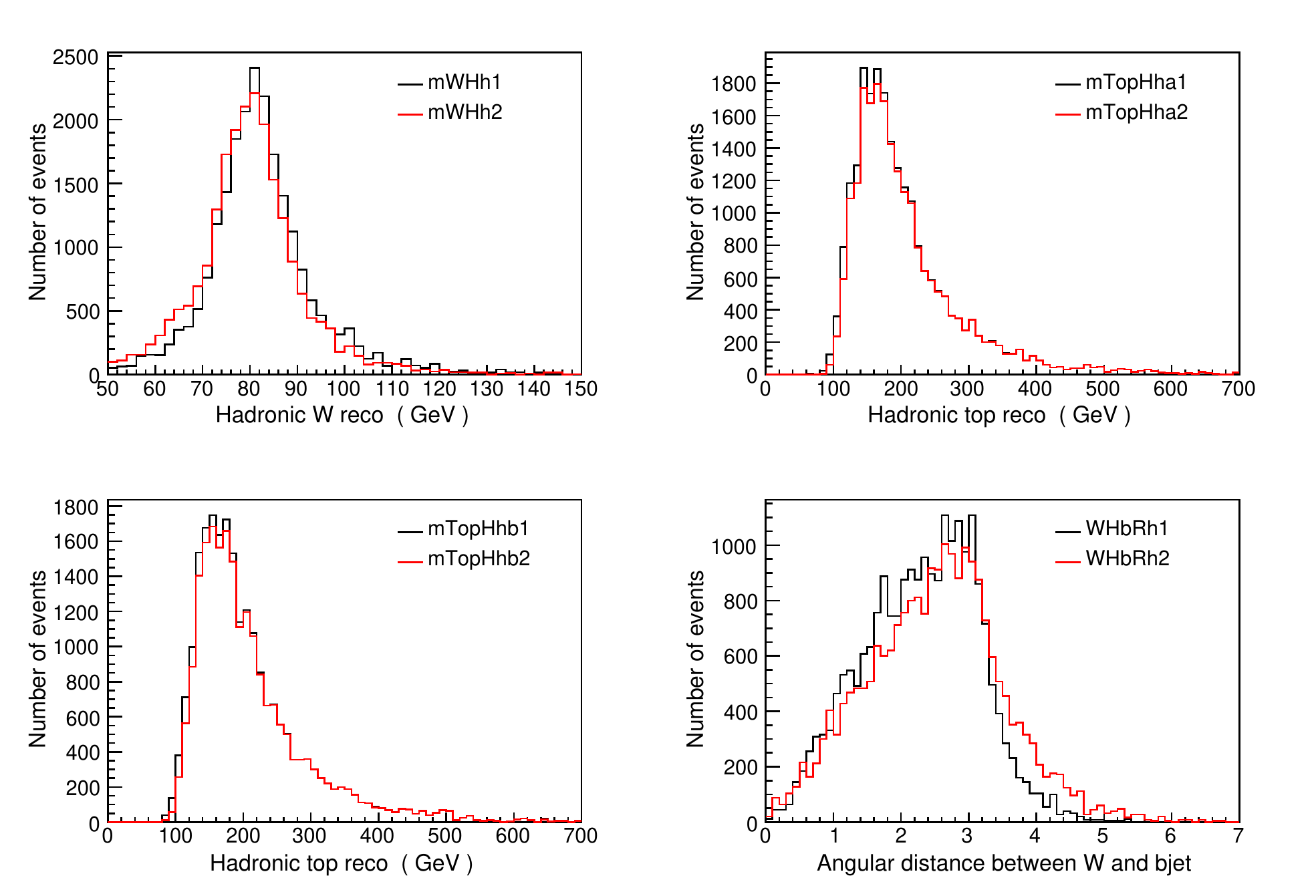}
\par\end{centering}
\caption{Results from fully hadronic top quark pair reconstruction example
with a single $\chi^2$ minimization of the top mass; Histograms with
names ending in 1 (2) are shown in black (red). \label{fig:single_chi2-tt}}
\end{figure}

The search described in the above subsection can also be implemented
in a single $\chi^{2}$ as defined in Equation~\ref{eq:chi2tt}:

\begin{equation}
\chi^{2}=\frac{(m_{b_{1}j_{1}j_{2}}-m_{b_{2}j_{3}j_{4}})^{2}}{\sigma_{\Delta m_{bjj}}^{2}}+\frac{(m_{j_{1}j_{2}}-m_{W}^{MC})^{2}}{\sigma_{\Delta m_{W}^{MC}}^{2}}+\frac{(m_{j_{3}j_{4}}-m_{W}^{MC})^{2}}{\sigma_{\Delta m_{W}^{MC}}^{2}}\quad.\label{eq:chi2tt}
\end{equation}
This single step reconstruction is used in the LHC hadronic top studies
\cite{topHadReco}. The \emph{CutLang} implementation of this case
and its results are given in Algorithm \ref{alg:wwtt-algo-1} and
Figure \ref{fig:single_chi2-tt}. Note that although about three times
slower on the same hardware, the single step approach yields about
1\% more events after all cuts as compared to the previous two step
approach. 

\begin{algorithm}
\caption{Top quark pair one step reconstruction example with $\chi^{2}$ search
and without $b$-tagging.\label{alg:wwtt-algo-1}}
\begin{lstlisting}

###### PARTICLE THRESHOLDS
minpte  = 15.0  # min pt of electrons 
minptm  = 15.0  # min pt of muons 
minptj  = 15.0  # min pt of jets
maxetae = 2.47  # max pseudorapidity of electrons  
maxetam = 2.5   # max pseudorapidity of muons 
maxetaj = 5.5   # max pseudorapidity of jets

TRGm = 0 #     muon Trigger Type: 0=dont trigger, 1=1st trigger (data) 2=2nd trigger (MC)
TRGe = 2 # electron Trigger Type: 0=dont trigger, 1=1st trigger (data) 2=2nd trigger (MC)

###### USER DEFINITIONS
def       "WH1 : JET_-1 JET_-1 "  # W boson of the first top
def       "WH2 : JET_-11 JET_-11 "  # W boson of the second top
def       "mWH1 : { WH1 }m "  # mass of W boson of the first top
def       "mWH2 : { WH2 }m "  # mass of @ boson of the second top
def       "mTopH1 : { WH1 JET_-2 }m "  # first top quark's mass
def       "mTopH2 : { WH2 JET_-4 }m "  # second top quark's mass
def       "WHbR1 : {WH1 , JET_-2 }dR "  # angular distance between W1 and b jet
def       "WHbR2 : {WH2 , JET_-4 }dR "  # angular distance between W2 and b jet

###### EVENT SELECTION
cmd      "ALL "           # to count all events
cmd      "nJET >= 6 "     # events with 6 or more jets
cmd      "FillHistos "
histo    "Basics "
cmd      "MET < 100 "     # fully hadronic events should have small MET
cmd      "(( mTopH1 - mTopH2 ) / 4.2  )^ 2 + mWH1 - 80.4 / 2.1 ^ 2 + mWH2 - 80.4 / 2.1 ^ 2 $\sim$= 0 " 
cmd      "FillHistos "
histo    "mWHh1 , Hadronic W reco (GeV), 50, 50, 150, mWH1 "
histo    "mWHh2 , Hadronic W reco (GeV), 50, 50, 150, mWH2 "
histo    "mTopHha1 , Hadronic top reco (GeV), 70, 0, 700, mTopH1 "
histo    "mTopHhb1 , Hadronic top reco (GeV), 70, 0, 700, mTopH2 "
histo    "WHbRh1 , Angular distance between W and bjet, 70, 0, 7, WHbR1 "
histo    "WHbRh2 , Angular distance between W and bjet, 70, 0, 7, WHbR2 "
cmd      "WHbR1 > 0.6 "
cmd      "WHbR2 > 0.6 "
cmd      "FillHistos "
histo    "mTopHha2, Hadronic top reco (GeV) after angular cut, 70, 0, 700, mTopH1 "
histo    "mTopHhb2, Hadronic top reco (GeV) after angular cut, 70, 0, 700, mTopH2 "
\end{lstlisting}
\end{algorithm}

\section{Processing speed\label{sec:Processing-speed}}

The speed at which a \emph{CutLang} based analysis processes events
was compared to that of a C$^{++}$ based, standalone and optimized
implementation of the same algorithm, keeping the same data access
functions, on a single core of a 3GHz, Intel-i7 CPU laptop from 2014.
For a simple Z boson reconstruction running on 500k events
in ATLAS Open Data format, the total computation time for the \emph{
CutLang} version was measured as 6.9s versus 6.3s for the standalone C$^{++}$ 
version, showing less than 10\% slowdown for the run-time interpretation 
of the human readable analysis description.  A more complete test was 
performed using the analysis with the fully hadronic $t\bar{t}$ events
as described in Section \ref{subsec:ana-ttbar}. Scanning the range
from 25k to 200k fully hadronic events from ATLAS Open Data and about
40k all inclusive events from CMS Open Data, the standalone C$^{++}$ version's 
speed was compared to that of the \emph{CutLang} analysis version.
The number of events and the complexity of the minimization algorithm
define the relative speeds of the two implementations. For the simpler
two step $\chi^{2}$ minimization, the C$^{++}$ version is 25\% faster
for smaller numbers of events (25k), reaching to about 50\% for 200k
events. For the more demanding single step version, the same comparison
ranges between 50\% to 65\%. These timings are deemed as affordable
costs as compared to the benefits provided by an interpreted, human
readable analysis description language and interpreter such as \emph{CutLang}, and
to the initial time taken by implementing even a simple analysis using
a generic computer language like C$^{++}$ or Python. For example, 
the one step and two step top quark reconstructions requiring one line and
two lines to implement in the \emph{CutLang} language take about 40 to 70
lines of standard analysis code in C$^{++}$. Finally, distributing the
large number of events onto a computing farm (e.g. using PROOF facilities
\cite{PROOF}) should help keeping the processing time taken by the \emph{CutLang} interpreter 
at a level comparable to the pure C$^{++}$ version of the same algorithm.

\section{Conclusions and outlook\label{sec:Conclusions-and-outlook}}

In this paper, a human readable analysis description language called
\emph{CutLang} and its runtime interpreter were presented. Although the current implementation should be seen as a proof-of-principle
version, it already addresses all main analysis needs such as variable, object and event selection definitions. \emph{CutLang} also
accommodates definition of multiple event selection regions and of multiple subsets leading to cleaned and tight physics objects.  Furthermore, the current version allows incorporation of user functions, the kind of selection
criteria that would be written in C$^{++}$ by the analyst, and require the re-compilation of the \emph{CutLang} source code, only once in the beginning.

An analysis description language like \emph{CutLang}, and its associated interpreter,  
provide a rapid development tool for complex analyses due to 
the ease with which ideas can be tested.  The unique runtime interpretation capability of \emph{CutLang} is a highly advantageous feature, especially during the analysis design phase, where object and event definitions are continuously modified to achieve an optimal selection.  The text-based language offers an easy way to write, understand and communicate an analysis.  
Such practicalities make \emph{CutLang} a suitable tool for many applications, however, \emph{CutLang} is primarily intended for analysis design, both in the experimental and phenomenological contexts.  It can be adapted for a full-fledged experimental analysis, or can easily be used for phenomenological analysis design, for example, for testing theoretical models, or devising new kinematic variables.  \emph{CutLang} can be very practical in sensitivity studies for future colliders.  Another important goal is to serve the physicists and physics enthusiasts working with the LHC open data to test new physics ideas.

\emph{CutLang} is currently being used in an ATLAS Exotics analysis, with, so far, positive feedback.  The junior researchers have praised the practicality brought by dispensing with the complexities of coding, whereas the senior researchers reported the ease of following the full scope of the analysis in complete detail.  
The persistency of an analysis also benefits from \emph{CutLang} since all threshold values and the analysis description are saved into the output ROOT file.

The experimental ATLAS analysis implementation naturally goes beyond object and event selection: it contains the necessary tools to define the signal and background samples, cross sections and  the systematic variations of about 200 different quantities.  Evidently, it is possible to completely define an analysis in a generic way, using human readable text files starting from the object selection up to the very last limit or discovery plot. Nevertheless, one should note that the sources of systematics variations, the methods to access those and the statistical analysis tools are usually experiment specific: no generic software tool can be used as an out of the box solution unless some serious effort is spent in defining various accords in those areas. 

A graphical user interface (GUI) for \emph{CutLang}, based on ROOT's graphics
libraries, is also under development. The GUI will allow the editing of the analysis description file, running
the analysis on a sample with appropriate command line parameters
and quickly displaying the output histograms. Since the design of \emph{CutLang}
preceded the LHADA proposal, \emph{CutLang} currently differs in syntax from 
that of LHADA. However, \emph{CutLang} fully embraces the LHADA
principles, chief among which is human readability and correctness.
A \emph{CutLang}-LHADA converter will also be available to bridge the gap
between the two.  In the future, when LHADA emerges as a mature language, a common ground between the two syntaxes can be found. The authors encourage experimentation with \emph{CutLang} using a variety of analyses in order to
help improve its internal algorithms and make the software more physicist friendly.

\section*{Acknowledgements}

The authors would like to H. B. Prosper for a careful reading of this
manuscript and useful suggestions, and E. Ozcan for an early implementation of the text file reader
code. The work of S.S. is supported by the National Research Foundation
of Korea (NRF), funded by the Ministry of Science \& ICT under contract
NRF-2008-00460 and by the U.S. Department of Energy through the Distinguished
Researcher Program from the Fermilab LHC Physics Center. G.U. would
like to dedicate this work to the ever enlightening memory of his
language and literature teacher Michel Tagan.

\section*{Bibliography}
\bibliographystyle{elsarticle-num} 
\bibliography{bibliography.bib}

\begin{thebibliography}{10}
\bibitem{showAllMacro}
The initial version of this ROOT macro was implemented by E. Ergenlik.
\end{thebibliography}
\pagebreak

\appendix

\renewcommand*{\appendixname}{}

\section{The \emph{CutLang} user manual\label{sec:userman}}
\begin{center}
V2.1 / May 2018 - For users with some experience in ROOT )
\par\end{center}

\subsection{How to obtain and build the \emph{CutLang} interpreter:}

The \emph{CutLang} interpreter package can be run on any standard Unix, Linux, or OSX machine which has an installation of the ROOT package.  Basic
knowledge of terminal operations such as text file editing and moving
files around is also needed in addition to knowledge of how to use the \emph{CutLang}
interpreter. Some knowledge on ROOT macros is helpful for manipulating histograms,
but is not essential for basic analysis operations.

The latest version of the \emph{CutLang} interpreter package can be downloaded
from HepForge at 
\begin{lstlisting}
http://cutlang.hepforge.org
\end{lstlisting}
The downloaded file should be opened with
\begin{lstlisting}
tar -xzf cutlang-VXX.YY.ZZ.tgz
\end{lstlisting}
which will automatically create the \texttt{CutLang} directory. The interpreter is
built using the commands
\begin{lstlisting}
cd CutLang/CLA
make
\end{lstlisting}
The interpreter should be executed from the \texttt{CutLang/runs} subdirectory. 

\subsection{How to run a \emph{CutLang} analysis }

Typically, running an analysis with \emph{CutLang} requires the following minimal set of files
\begin{itemize}
\item The text based analysis description file (examples can be found under directory \texttt{runs}, which can be used as examples to create the user's analysis description files. Default file name is \texttt{CLA.ini}, however the file name can be specified at the command line.) 
\item The shell script file \texttt{CLA.sh} : The script that executes the \emph{CutLang} analysis (see Step 4 below).
This script takes two mandatory arguments:  i) the name of the ROOT file to be analyzed, e.g., \texttt{cms-opendata-ttbar.root}, which is to be downloaded from the relevant source or generated by the user externally; and ii) the input data format.  The already existing input data formats are LVL0 (\emph{CutLang} default format), ATLASOD (ATLAS open data), CMSOD (CMS open data), Delphes, FCC, LHCO.  New input data formats can be added as to be described in Appendix~\ref{sec:newinputformat}.
The other arguments, specified below, are optional:
\begin{lyxlist}{00.00.0000}
\item [{\texttt{-e|-{}-events}}] the number of events to be processed.
By default all events (represented by 0) are processed. 
\item [{\texttt{-i|-{}-inifile}}] the analysis description file to be processed.
By default \texttt{CLA.ini} is processed.
\item [{\texttt{-v|-{}-verbose}}] the verbosity event count. By default after
every 1000 events, the current event count is written on the screen. This option allows to change the default number of events reported, e.g. as \texttt{--verbose 5000}. 
\item [{\texttt{-h|-{}-help}}] displays these explanations as reminders.
\end{lyxlist}

\item \texttt{histoOut-NAME.root} output file produced in Step 8 where
NAME is the name of the analysis description file. If no name is specified, the default value of \texttt{histoOut-CLA.root} is used.
In case of multiple signal regions, each region will have its own directory inside the output file marked with \texttt{BP\_i} where
\texttt{i} is an index number.

\item The ROOT file containing events to be analyzed (A small sample of $t\bar{t}$ events can be downloaded from \href{http://cutlang.hepforge.org}{http://cutlang.hepforge.org}
for a quick start). Multiple comma separated input files and their
paths can be specified at the command line as; e.g.,
\begin{quote}
\texttt{./CLA.sh
../roots/atlas1ttbar.root,../roots/atlas2ttbar.root ATLASOD}
\end{quote}
\end{itemize}

The steps below should be followed for running an analysis in \emph{CutLang}.
\begin{enumerate}
\setcounter{enumi}{-1}
\item Open a terminal
\item Go to directory \texttt{CutLang}/runs
\item Edit the analysis description file, e.g. \texttt{CLA.ini}, as needed 
\item Go back to the same terminal
\item Execute the analysis description edited in Step 2 using the following command:
\begin{lstlisting}
./CLA.sh [input ROOT file(s)] [data format] [-i youranalysis.ini] 
         [-e number_of_events_to_process] [-v verbosity]
\end{lstlisting}
An example would be
\begin{lstlisting}
./CLA.sh cms-opendata-ttbar.root CMSOD -i CLA.ini -e 10000 
\end{lstlisting}

\item \emph{CutLang} outputs the analysis evaluation results to the terminal.
\begin{itemize}
\item If the analysis description file is syntactically correct, the following message should
appear on the screen: `` \texttt{End
of analysis initialization}" , in which case all is well and proceed to Step 6.
\item If there are any errors in the analysis description, \emph{CutLang}
notifies the user of the unknown parameter(s) as `` \texttt{\textbf{UFO(s)}}" .
Go back to Step 2, verify and correct the ``\texttt{ini}" file.
\end{itemize}
\item \texttt{CLA} lists messages every 1000 processed events until
it reaches the end of the ROOT file. 
\item An efficiency table for the analysis is displayed on the screen. 
\item \texttt{CLA} displays the message: `` \texttt{saving\dots finished}"
at the end of the analysis. The output ROOT file is saved under the same directory.
The file name will be \texttt{histoOut-NAME.root}, where
NAME is the name of the analysis description (\texttt{ini}) file. If no name
is specified, the default value of \texttt{histoOut-CLA.root} is used.
If users wish to keep output from a previous run, the output file
should be renamed. Otherwise, the \texttt{CLA.sh} will overwrite the
output file. 
In case of multiple signal regions, each region will have its own
directory inside the output file marked with \texttt{BP\_i} where \texttt{i} is an index number.  
\end{enumerate}

\subsection{How to prepare a \emph{CutLang} analysis description file }

The analysis description file (e.g. \texttt{CLA.ini}) contains three sections: 
\begin{itemize}
\item {\bf Object thresholds:} This mandatory section contains the
$\eta$ and $p_{T}$ threshold values for a particle to be accepted.
\item {\bf User definitions:} This is a non-mandatory section containing
user definitions starting with keyword ``\texttt{def}" 
for new composite particles and variables. These definitions can be used to
create shorthand names for otherwise long expressions for
later use.  This section can also contain the derived objects sets, such as
cleaned or tighter objects which might be used in the event selection. The object definitions
should start with the with keyword ``\texttt{obj}". 
\item {\bf Event selection:} This section is mandatory and consists
of lines starting with keyword `` \texttt{cmd}",
which define event operations or selection criteria. The section also consists
of lines starting with `` \texttt{histo}",
which signify the histogram definitions.
\end{itemize}

\begin{center}
\em{Common rules for all sections}
\par\end{center}
\begin{itemize}
\item All lines start with one of the \texttt{def}, \texttt{cmd}, or \texttt{histo}
keywords. No space should exist before the keywords. Note that there
are no keywords for the object thresholds section.
\item An indefinite amount of space is allowed between the keyword and the
command/description.
\item Every command/description must be enclosed within double quotation
marks, and there should be a space before the ending quotation mark,
e.g. \texttt{"mLL : \{ LEP\_1 LEP\_0 \}m "}, or \texttt{"mLL [] 70 120 "}.
\item There is no upper limit for the number of lines.
\item At least one space must be left before and after each term; including
operands, numbers.
\item All units in \emph{CutLang} are either GeV or radians ($c=1$, therefore
mass, momentum, energy are all in GeV)
\item All variable, function, and particle names are case sensitive.
\end{itemize}

\begin{center}
\em{Additional rules in editing the analysis description file}
\par\end{center}

\begin{itemize}
\item ``$\sim=$" and ``$!=$" cannot be combined with any other
operator or function. For example:

`` \texttt{nJET} $>=$ 6 AND \texttt{nBJET} $>=$ 2 "
\textbf{\emph{OK}}

`` \texttt{mTopb} $\sim=$175 " \emph{}\textbf{\emph{OK}}

`` \texttt{mTopb} $\sim=$ 175 AND \texttt{nBJET} $>=$ 2 " \textbf{\emph{not OK}}

`` \texttt{mTopb} $\sim=$ 175 AND \texttt{mTopb2}
$\sim=$175 " \textbf{\emph{not OK}}

\item Any expression after a ``\#" is considered
a comment and ignored. Consequently, lines can be skipped by 
commenting them out.  
\item  Names of the user defined composite particles and variables
must be unique. They cannot be redefined within the same analysis.
\item  ``\{ \}" are used for stating properties of particles, e.g. mass, charge, etc. One can add as
many particles as required in the term within the curly braces. 
\item  The order of particles in a particle combination is immaterial; i.e. : \texttt{LEP\_0 LEP\_1} is 
functionally identical to \texttt{LEP\_1 LEP\_0}.
\item  All definitions in the \texttt{def} section should be ordered with the lower case and upper case reverse alphabetical order respectively.
\end{itemize}

\begin{center}
\em{A note on the trigger scope}
\par\end{center}
One should note that only electron and muon triggers are implemented in this version. (The tau channel is not available in this version.)  Wherever the term ``lepton" or the abbreviation ``LEP" is  used, this refers to either an electron or a muon depending on the selected trigger. 
A trigger value of 0 deselects that channel, 1 treats the input file as data (i.e. no event weights are applied), and finally 2 ensures the application of all the relevant weights such as Monte Carlo weights, pileup weights,  vertex weights and b-tagging weights. For a given analysis run, only a single lepton type can be triggered, while the other type has to be set to 0.
 \\

\subsubsection{An example analysis description file with multiple regions}

\begin{lstlisting}
###### OBJECT THRESHOLDS
minpte = 15.0 # min pt of electrons
minptm = 15.0 # min pt of muons
minptj = 15.0 # min pt of jets
maxetae = 2.47 # max pseudorapidity of electrons
maxetam = 2.5 # max pseudorapidity of muons
maxetaj = 5.5 # max pseudorapidity of jets
TRGm = 0 # muon Trigger Type: 0=dont trigger, 1=1st trigger (data) 2=2nd trigger (MC)
TRGe = 2 # electron Trigger Type: 0=dont trigger, 1=1st trigger (data) 2=2nd trigger (MC)

###### USER DEFINITIONS
def "WH1 : JET_-1 JET_-1 " # W boson of the first top
def "WH2 : JET_-11 JET_-11 " # W boson of the second top
def "mWH1 : { WH1 }m " # mass of W boson of the first top
def "mWH2 : { WH2 }m " # mass of @ boson of the second top
def "mTopH1 : { WH1 JET_-2 }m " # first top quark's mass
def "mTopH2 : { WH2 JET_-4 }m " # second top quark's mass

###### EVENT SELECTION
algo __preselection__
cmd "ALL " # to count all events
cmd "nJET >= 6 " # events with 6 or more jets
cmd "MET < 100 " # fully hadronic events should have small MET
#cmd "FillHistos "
#histo "Basics "

algo __teknik1__
__preselection__
cmd "mTopH1 - mTopH2 / 4.2 ^ 2 + mWH1 - 80.4 / 2.1 ^ 2 + mWH2 - 80.4 / 2.1 ^ 2 $\sim$= 0 "
cmd "FillHistos "
histo "mWHh1 , Hadronic W reco (GeV), 50, 50, 150, mWH1 "
histo "mWHh2 , Hadronic W reco (GeV), 50, 50, 150, mWH2 "
histo "mTopHha1 , Hadronic top reco (GeV), 70, 0, 700, mTopH1 "
histo "mTopHhb1 , Hadronic top reco (GeV), 70, 0, 700, mTopH2 "

algo __teknik2__
__preselection__
cmd "mWH1 - 80.4 / 2.1 ^ 2 + mWH2 - 80.4 / 2.1 ^ 2  $\sim$= 0 " # 2 WHads
cmd "mTopH1 - mTopH2 / 4.2 ^ 2 ~= 0 "
cmd "FillHistos "
histo "mWHh1 , Hadronic W reco (GeV), 50, 50, 150, mWH1 "
histo "mWHh2 , Hadronic W reco (GeV), 50, 50, 150, mWH2 "
histo "mTopHha1 , Hadronic top reco (GeV), 70, 0, 700, mTopH1 "
histo "mTopHhb1 , Hadronic top reco (GeV), 70, 0, 700, mTopH2 "
\end{lstlisting}

\vspace{1cm}

\subsection{How to view a \emph{CutLang} analysis output}

There are two ways to view the contents of the \emph{CutLang} output ROOT file: 
\begin{enumerate}
\item Open it using ROOT : \texttt{root.exe histoOut-CLA.root;} and launch
a \texttt{TBrowser}
\item Run the default macro: \texttt{./showall.sh} :
\begin{lstlisting}
./showall.sh [regionID] [histofileName]
\end{lstlisting}
This shows the results for the region {\texttt regionID}.  Default values are {\texttt 1} and {\texttt histoOut-CLA.root}.
\end{enumerate}

The following few lines show the typical beginning of an output ROOT file.  Note that the user definitions, and the cutflow all in \emph{CutLang} format are reproduced for the reader's convenience.  Moreover, the object definition threshold values are stored in \texttt{TParameter} variables.  The efficiency histograms which are always automatically booked and filled are also shown.

\begin{lstlisting}
root [2] .ls
TDirectoryFile*		BP_2	BP_2
 KEY: TText	CLA2defs;1	
WH1 : JET_-1 JET_-1  
WH2 : JET_-11 JET_-11  
mWH1 : { WH1 }m  
mWH2 : { WH2 }m  
mTopH1 : { WH1 JET_-2 }m  
mTopH2 : { WH2 JET_-4 }m  
WHbR1 : {WH1 , JET_-2 }dR  
WHbR2 : {WH2 , JET_-4 }dR  
Wchi2 : { WH1 }m - 80.4 / 2.1 ^ 2 + { WH2 }m - 80.4 / 2.1 ^ 2  
topchi2 : mTopH1 - mTopH2 / 4.2 ^ 2  

 KEY: TText	CLA2cuts;1	
cmd1 : ALL 
cmd2 : nJET >= 6 
cmd3 : MET < 100 
cmd4 : topchi2 + Wchi2 $\sim$= 0 
cmd5 : FillHistos 
cmd6 : WHbR1 > 0.6 
cmd7 : WHbR2 > 0.6 
cmd8 : FillHistos 

 KEY: TParameter<double>	minpte;1	
 KEY: TParameter<double>	maxetae;1	
 KEY: TParameter<double>	minptm;1	
 KEY: TParameter<double>	maxetam;1	
 KEY: TParameter<double>	minptj;1	
 KEY: TParameter<double>	maxetaj;1	
 KEY: TParameter<double>	TRGe;1	
 KEY: TParameter<double>	TRGm;1	
 KEY: TH1F	eff;1	selection efficiencies 
\end{lstlisting}

\subsection{How to interface a new input data format to the \emph{CutLang} interpreter \label{sec:newinputformat}}

This section describes how to build the interface between a new data file format represented as a 
flat ntuple and the standard types used by the \emph{CutLang} interpreter. This is one aspect of the current version of
\emph{CutLang} that requires some coding expertise.  \emph{CutLang} uses ROOT's \texttt{MakeClass} for this purpose.

\begin{itemize}
\item Obtain a sample ROOT ntuple file containing the new data format and load into ROOT (e.g., using \texttt{TFile f("myfile.root")})
\item Call the ROOT \texttt{MakeClass} command on the relevant tree, specifying a class name
\begin{verbatim}
tree->MakeClass("NewFormatName");
\end{verbatim}
\item Move the resulting header file (\texttt{NewFormatName.h} ) into the
\texttt{analysis\_core} subdirectory, and is include it in the main code \texttt{CLA.C} 
\item Move the resulting implementation macro (\texttt{NewFormatName.C}) into the
\texttt{CutLang/CLA} directory, and include the following required headers in it:
\begin{lstlisting}
#include "lhco.h"
#include <TH2.h>}
#include <TStyle.h>}
#include <TCanvas.h>}
#include <signal.h>}
#include "dbx_electron.h"
#include "dbx_muon.h"
#include "dbx_jet.h"
#include "dbx_a.h"
#include "DBXNtuple.h"
#include "analysis_core.h"
#include "AnalysisController.h"
\end{lstlisting}

\item In the event loop, the input data must be transferred to the standard \emph{CutLamg} types, e.g., the electron, muon, photon
and jet particle vectors, without forgetting any available event-wide information like RunNumber, EventNumber etc. An example conversion for the \texttt{LHCO} format is:
\begin{lstlisting} 
TLorentzVector alv; dbxMuon *adbxm; vector<dbxMuon> muons; 
for (unsigned int i=0; i<Muon_; i++) {
  alv.SetPtEtaPhiM(Muon_PT[i], Muon_Eta[i], Muon_Phi[i], (105.658/1E3)); // all in GeV
  adbxm= new dbxMuon(alv);
  adbxm->setCharge(Muon_Charge[i] );
  adbxm->setEtCone(Muon_ETiso[i] );
  adbxm->setPtCone(Muon_PTiso[i] );
  adbxm->setParticleIndx(i);
  muons.push_back(*adbxm);
  delete adbxm;
  }
\end{/}
\item Modify the end of the macro to be as follows:
\begin{lstlisting}
AnalysisObjects a0={muons, electrons, photons, jets, met, anevt};
aCtrl.RunTasks(a0);
} // end of event loop
aCtrl.Finalize();
} // end of Loop function
\end{lstlisting}
\end{itemize}

\subsection{How to add user functions \label{sec:userfuncdetail}}
As discussed in Section~\ref{sec:userfunc}, it is possible to add (or delete) user functions to \emph{CutLang} for computation of complex variables.  This can be done using the \texttt{scripts/adduserfunction.py} script. To add a function, run
\begin{lstlisting}
python adduserfunction.py <functionname>
\end{lstlisting}
This creates the function header \texttt{analysis\_core/dbx\_<functionname>.h} and adds the function into \texttt{analysis\_core/dbxCut.cpp}.  The content of an example user function \texttt{dbx\_userfunc1.h} is shown below:

\begin{lstlisting}
#ifndef DBX_USERFUNC1_H
#define DBX_USERFUNC1_H

class dbxCutuserfunc1 : public dbxCut {
 public:
     dbxCutuserfunc1: dbxCut("}userfunc1"){}
     dbxCutuserfunc1(std::vector<int> ts, std::vector<int> is, int v )
                                 : dbxCut("}s(name)s",ts,is,v){}

      bool select(AnalysisObjects *ao){
          float result;
          result=calc(ao);
          return (Ccompare(result) ); 
      }
      float calc(AnalysisObjects *ao){
         float retval;

/* this is an example on how to calculate Meff in an external function.
// each particle given to this function is of  type getParticleType(jj)
// each particle given to this function has index getParticleIndex(jj)

float meff=0;
for (unsigned int jj=0; jj<2; jj++)  //Meff = MET + sum of all a particle type's all PTs
  switch (getParticleType(jj*2)){
    case 0: for (int ii=0; ii<ao->muos.size(); ii++) 
                 meff+=ao->muos[ii].lv().Pt();
                 break;
    case 1: for (int ii=0; ii<ao->eles.size(); ii++) 
                 meff+=ao->eles[ii].lv().Pt();
                 break;
    case 2: for (int ii=0; ii<ao->jets.size(); ii++) 
                 meff+=ao->jets[ii].lv().Pt();  //these are un-tagged jets
                 break;               
    case 3: for (int ii=0; ii<tagJets(ao,1).size(); ii++) 
                 meff+=tagJets(ao,1)[ii].lv().Pt(); //these are b-tagged jets
                 break;                
    case 4: for (int ii=0; ii<tagJets(ao,0).size(); ii++) 
                 meff+=tagJets(ao,0)[ii].lv().Pt();  //these are b-tag rejected jets
                 break;              
    case 7: meff+=ao->met.Mod();
                 break;
    case 8: for (int ii=0; ii<ao->gams.size(); ii++) 
                 meff+=ao->gams[ii].lv().Pt();
                 break;
}
retval=meff;
*/


         // ***********************************
         // Write your own code here
         // ***********************************
         
         return retval;
}
private:
};
\end{lstlisting}

If for some reason, the user function needs to be deleted, this can also be done safely with the same script using
\begin{lstlisting}
python adduserfunction.py --delete <functionname>
\end{lstlisting}

\section{Two example analyses in \emph{CutLang} \label{sec:furtherexamples}}
In the following, we present implementations of two real life analyses written using \emph{CutLang}.  The analyses are taken from a recent study comparing public recasting tools done within the context of Les Houches PhysTeV 2017 proceedings~\cite{Brooijmans:2018xbu} (see Section 21).  The first example is an ATLAS exotic monophoton search~\cite{Aaboud:2017dor} with detailed object definitions and multiple event selection regions defined by different missing transverse momentum thresholds.  The second example is an ATLAS SUSY search in the jets and missing transverse momentum final state~\cite{Aaboud:2016zdn}, which also has a detailed object selection and multiple event selection regions defined by several complex selection variables.  These analyses were run using \emph{CutLang} on signal events generated for the Les Houches study (as described in~\cite{Brooijmans:2018xbu}), and simulated privately using Delphes.  The results obtained are very close to those presented in~\cite{Brooijmans:2018xbu}, and exactly the same with those obtained from a private comparison with a recent LHADA interpreter called \emph{lhata2tnm} (described in Section 27 of~\cite{Brooijmans:2018xbu}).  For the latter comparison, the same Delphes samples were used by \emph{CutLang} and \emph{lhada2tnm}.

%\vspace{1cm}

\newpage

\noindent {\bf Example 1: }ATLAS exotic monophoton analysis
\vspace{-0.3cm}
\begin{lstlisting}[
    basicstyle=\small, %or \small or \footnotesize etc.
]

# info analysis
#   experiment ATLAS
#   id EXOT-2016-32
#   publication Eur.Phys.J. C77 (2017) no.6, 393
#   sqrtS 13.0
#   lumi 36.1
#   arXiv 1704.03848
#   hepdata https://www.hepdata.net/record/ins1591328
#   doi 10.1140/epjc/s10052-017-4965-8

######## GENERIC OBJECT THRESHOLDS
minptp  = 10.0  # min pt of photons
minpte  =  7.0  # min pt of electrons
minptm  =  6.0  # min pt of muons
minptj  = 20.0  # min pt of jets

maxetap = 2.37  # max pseudorapidity of photons
maxetae = 2.47  # max pseudorapidity of electrons
maxetam = 2.70  # max pseudorapidity of muons
maxetaj = 4.50  # max pseudorapidity of jets

######## OBJECT SELECTION
obj  "JETclean : JET "
cmd  "{ JET_ , ELE_ }dR >= 0.2 "

obj  "ELEclean : ELE "
cmd  "{ ELE_ , JETclean_ }dR >= 0.4 "

obj  "MUOclean : MUO "
cmd  "{ MUO_ , JETclean_ }dR >= 0.4 "

obj  "PHOtight : PHO "
cmd  "{ PHO_ }AbsEta ][  1.37 1.52 "

obj  "JETsr : JETclean "
cmd  "{ JETclean_ }Pt > 30 "
cmd  "{ JETclean_ , PHOtight_ }dR >= 0.4 "
cmd  "{ JETclean_ , METLV_0 }dPhi >= 0.4 "

######## EVENT SELECTION
algo __preselection__
cmd   "ALL "                          # to count all events
cmd   " nPHOtight >= 0 "              # events with 1 or more tight photons
cmd   "{ PHOtight_0 }Pt > 150 "         # select photons[0].PT > 150
cmd   "{ PHOtight_0 , METLV_0 }dPhi > 0.4 "         # select isolated photons
cmd   " MET  HT ^ 0.5 / > 8.5 "           # select METoverSqrtSumET > 8.5
cmd   " nJETsr <= 1 "
cmd   "nJETsr == 0 ? ALL : { JETsr_0 , METLV_0 }dPhi > 0.4 "   # select dphi(jetsSR.Phi, MET.Phi) > 0.4
cmd   "  nMUOclean == 0 "             # select cleanmuons.size == 0
cmd   "  nELEclean == 0 "             # select cleanelectrons.size == 0


# Inclusive search regions
algo __SRI1__
__preselection__
cmd      "MET > 150 "
algo __SRI2__
__preselection__
cmd      "MET > 225 "
algo __SRI3__
__preselection__
cmd      "MET > 300 "

# Exclusive search regions
algo __SRE1__
__preselection__
cmd      "MET [] 150 225 "
algo __SRE2__
__preselection__
cmd      "MET [] 225 300 "
\end{lstlisting}

\vspace{0.5cm}

\noindent {\bf Example 2: }ATLAS SUSY JetMET analysis

\begin{lstlisting} [
    basicstyle=\small, %or \small or \footnotesize etc.
]
# info analysis
#  experiment ATLAS
#  id SUSY-2013-15
#  publication Eur. Phys. J. C(2016) 76: 392
#  sqrtS 13.0
#  lumi 3.2
#  arXiv 1605.03814
#  hepdata http://hepdata.cedar.ac.uk/view/ins1304456
#  doi 10.1140/epjc/s10052-016-4184-8

######## GENERIC OBJECT THRESHOLDS
minptp  = 10.0  # min pt of photons
minpte  = 10.0  # min pt of electrons
minptm  = 10.0  # min pt of muons
minptj  = 20.0  # min pt of jets

maxetap = 2.37  # max pseudorapidity of photons
maxetae = 2.47  # max pseudorapidity of electrons
maxetam = 2.70  # max pseudorapidity of muons
maxetaj = 2.80  # max pseudorapidity of jets

######## USER DEFINITIONS
def "Meff : MET + HT "  #Meff is simple
def "JM0 : { JETsr_0 , METLV_0 }dPhi "
def "JM1 : { JETsr_1 , METLV_0 }dPhi "
def "JM2 : { JETsr_2 , METLV_0 }dPhi "
def "Meff4j : MET + { JETsr_0 }Pt + { JETsr_1 }Pt + { JETsr_2 }Pt + { JETsr_3 }Pt "
def "Meff5j : Meff4j + { JETsr_4 }Pt "
def "Meff6j : Meff5j + { JETsr_5 }Pt "

######## OBJECT SELECTION
obj  "JETclean : JET "
cmd  "{ JET_ , ELE_ }dR >= 0.2 "

obj  "MUOclean : MUO "
cmd  "{ MUO_ , JETclean_ }dR >= 0.4 "
cmd  "{ MUO_ }IsolationRhoCorr < 0.1"

obj  "ELEclean : ELE "
cmd  "{ ELE_ , JETclean_ }dR >= 0.4 "

obj  "ELEveryclean : ELE "
cmd  "{ ELE_ , JETclean_ }dR >= 0.4 "

obj  "JETsr : JETclean "
cmd  "{ JETclean_ }Pt > 50 "

######## EVENT SELECTION
algo __preselection__
cmd   "ALL "                          # to count all events
cmd   "MET > 200 "
#cmd   "nPHOtight >= 0 "
cmd   "nMUOclean == 0 "         # Reject evt if there is a muon with pT > 10
#cmd   "nMUOclean == 0 ? ALL : { MUOclean_0 }Pt < 10 "         # Reject evt if there is a muon with pT > 10
cmd   "nELEveryclean == 0 "         # Reject evt if there is a muon with pT > 10
#cmd   "nELEveryclean == 0 ? ALL : { ELEveryclean_0 }Pt < 10 "    # Reject evt if there is an electron with pT > 10
cmd   "nJETsr > 0 "

algo __2jt__
__preselection__
cmd   "nJETsr >= 1 "
cmd   "{ JETsr_0 }Pt > 200 "
cmd   "nJETsr >= 2 "
cmd   "Ex1 ( JETsr_ ) > 0.8  "
#cmd   "nJETsr == 2 ?  JM0 - JM1 < 0   ?  JM0 > 0.8 : JM1 > 0.8 : ALL "
#cmd   "JM0 - JM1 < 0 ? JM0 - JM2 < 0 ? JM0 > 0.8 : JM2 > 0.8 : JM1 - JM2 < 0 ? JM1 > 0.8 : JM2 > 0.8 "
cmd   "{ JETsr_1 }Pt > 200 "
cmd   "MET / HT ^ 0.5  ( JETsr_ ) > 20 "
cmd   "Meff ( JETsr_ ) > 2000 "

algo __2jm__
__preselection__
cmd   " nJETsr >= 1 "
cmd   "{ JETsr_0 }Pt > 300 "
cmd   " nJETsr >= 2 "
cmd   " Ex1 ( JETsr_ ) > 0.4  "
cmd   "{ JETsr_1 }Pt > 50 "
cmd   " MET / HT ^  0.5 ( JETsr_ )  > 15 "
cmd   " Meff ( JETsr_ ) > 1600 "

algo __2jl__
__preselection__
cmd   "nJETsr >= 1 "
cmd   "{ JETsr_0 }Pt > 200 "
cmd   "nJETsr >= 2 "
cmd   "Ex1 ( JETsr_ ) > 0.8  "
cmd   "{ JETsr_1 }Pt > 200 "
cmd   "MET / HT ^ 0.5  ( JETsr_ ) > 15 "  # make sure we use JETsr in HT
cmd   "Meff  ( JETsr_ ) > 1200 "

algo __4jt__
__preselection__
cmd   " nJETsr >= 4 "
cmd   "{ JETsr_0 }Pt > 200 "
cmd   "{ JETsr_1 }Pt > 100 "
cmd   "{ JETsr_2 }Pt > 100 "
cmd   "{ JETsr_3 }Pt > 100 "
cmd   "Ex1 ( JETsr_ ) > 0.4  "
cmd   "{ JETsr_-1 , METLV_0 }dPhi $\sim$= 0.0 "
cmd   "{ JETsr_-1 , METLV_0 }dPhi > 0.2 "
#cmd   "aplanarity ( JETsr_ ) > 0.04 "
cmd   "MET / Meff4j ( JETsr_ ) > 0.2 "
cmd   "Meff (JETsr_ ) > 2200 "

algo __5j__
__preselection__
cmd   " nJETsr >= 5 "
cmd   "{ JETsr_0 }Pt > 200 "
cmd   "{ JETsr_1 }Pt > 100 "
cmd   "{ JETsr_2 }Pt > 100 "
cmd   "{ JETsr_3 }Pt > 100 "
cmd   "{ JETsr_4 }Pt > 50 "
cmd   "Ex1 ( JETsr_ ) > 0.4  "
cmd   "{ JETsr_-1 , METLV_0 }dPhi $\sim$= 0.0 "
cmd   "{ JETsr_-1 , METLV_0 }dPhi > 0.2 "
#cmd   "aplanarity ( JETsr_ ) > 0.04 "
cmd   "MET / Meff5j ( JETsr_ ) > 0.25 "
cmd   "Meff ( JETsr_ ) > 1600 "

algo __6jm__
__preselection__
cmd   " nJETsr >= 6 "
cmd   "{ JETsr_0 }Pt > 200 "
cmd   "{ JETsr_1 }Pt > 100 "
cmd   "{ JETsr_2 }Pt > 100 "
cmd   "{ JETsr_3 }Pt > 100 "
cmd   "{ JETsr_4 }Pt > 50 "
cmd   "{ JETsr_5 }Pt > 50 "
cmd   "Ex1 ( JETsr_ ) > 0.4  "
cmd   "{ JETsr_-1 , METLV_0 }dPhi $\sim$= 0.0 "
cmd   "{ JETsr_-1 , METLV_0 }dPhi > 0.2 "
#cmd   "aplanarity ( JETsr_ ) > 0.04 "
cmd   "MET / Meff6j ( JETsr_ )  > 0.25 "
cmd   "Meff  ( JETsr_ ) > 1600 "

algo __6jt__
__preselection__
cmd   " nJETsr >= 6 "
cmd   "{ JETsr_0 }Pt > 200 "
cmd   "{ JETsr_1 }Pt > 100 "
cmd   "{ JETsr_2 }Pt > 100 "
cmd   "{ JETsr_3 }Pt > 100 "
cmd   "{ JETsr_4 }Pt > 50 "
cmd   "{ JETsr_5 }Pt > 50 "
cmd   "Ex1 ( JETsr_ ) > 0.4 "
cmd   "{ JETsr_-1 , METLV_0 }dPhi $\sim$= 0.0 "
cmd   "{ JETsr_-1 , METLV_0 }dPhi > 0.2 "
#cmd   "aplanarity ( JETsr_ ) > 0.04 "
cmd   "MET / Meff6j ( JETsr_ )  > 0.2 "
cmd   "Meff ( JETsr_ ) > 2000 "
\end{lstlisting}

\vspace{0.5cm}

Note that in the second example, the region defined by the algorithm called {\tt 2jt}  contains an external user function called
{\tt Ex1}. This external function finds the smallest polar angular distance between MET and the first three jets. When there are more jets, only the first three are taken into account, while 
 when there are two jets, only those available two jets are considered. The same function can also be implemented using the ternary functions and comparison operators all available in {\em CutLang}.
 The  two commented out lines just below the {\tt Ex1} function show how to do that same computation using the ternary functions and comparison operators.

\end{document}